\documentclass[aps,prd,twocolumn,superscriptaddress,preprintnumbers,nofootinbib,longbibliography,floatfix]{revtex4-1}

\usepackage{graphicx}
\usepackage{dcolumn}
\usepackage{bm}
\usepackage{ragged2e}
\usepackage[hidelinks]{hyperref}
\usepackage[mathlines]{lineno}
\usepackage[caption=false]{subfig}
\usepackage{colortbl}
\usepackage{enumitem}
\usepackage{amsmath}

\hypersetup{breaklinks=true}
\newcommand{\eg}{$e.g.$}

\renewcommand{\email}[1]{\href{mailto:#1}{\texttt{#1}}}
\newcolumntype{P}[1]{>{\centering\arraybackslash}p{#1}}

\begin{document}

\preprint{UCI-TR-1234-56}
\preprint{LLNL-JRNL-819142}

\title{Feasibility of Correlated Extensive Air Shower Detection\texorpdfstring{\\with a Distributed Cosmic Ray Network}{}}

\author{Eric Albin}\thanks{Corresponding author, albin3@llnl.gov}
\affiliation{Department of Physics and Astronomy, UC Irvine, Irvine, CA 92627}
\affiliation{Lawrence Livermore National Laboratory, Livermore, CA 94550}

\author{Daniel Whiteson}
\affiliation{Department of Physics and Astronomy, UC Irvine, Irvine, CA 92627}


\begin{abstract}
We explore the sensitivity offered by a global network of cosmic ray detectors to a novel, unobserved phenomena: widely separated simultaneous extended air showers.  
Existing localized observatories work independently to observe individual showers, offering insight into the source and nature of ultra-high energy cosmic rays. 
However no current observatory is large enough to provide sensitivity to anticipated processes such as the GZ effect or potential new physics that generate simultaneous air showers separated by hundreds to thousands of kilometers. 
A global network of consumer electronics (the CRAYFIS experiment), may provide a novel opportunity for observation of such phenomena. Two user scenarios are explored. In the first, with maximal user adoption, we find that statistically significant discoveries of spatially-separated but coincident showers are possible within a couple years. In the second, more pratical adoption model with $10^6$ active devices, we find a worldwide CRAYFIS to be sensitive to novel ``burst'' phenomena where many simultaneous EASs occur at once. 
\end{abstract}

\maketitle

\tableofcontents


\section{Introduction}
\label{sec:introduction}

The history of experimental astrophysics offers many examples of surprising discoveries that come unexpectedly following the introduction of new observational technologies. Well known examples include the famous discovery of the CMB~\cite{CMB} and the more recent observation of the Fermi bubbles~\cite{Fermi_Bubble_1, Fermi_Bubble_2}. A globally-distributed network of cosmic ray sensors would be a novel instrument for the exploration of cosmic rays~\cite{CREDO:2020}, with the potential to reveal unexpected or previously-unobserved planet-scale phenomena such as  widely-separated simultaneous extensive air showers. 

Comparatively smaller-scale and higher-cost campaigns to detect simultaneous showers have been ongoing for the past 50 years~\cite{Fegan:1984, Carrel:1994, Kitamura:1997, Unno:1997, Ochi:1999, ICRC:1998, ICRC:1999, ICRC:2001, Schools}.   A truly global network of dedicated devices would be prohibitively expensive, but concepts have been developed for the repurposing of consumer smartphones to detect extensive air showers~\cite{crayfis, CREDO:2020, DECO:2017}.   Such a network may have the capability to observe widely separated correlated air showers, or other unexpected phenomena which require global scale.

Several theoretical mechanisms for global phenomena have been described in the literature. A leading example is the Gerasimova-Zatsepin (GZ) effect---simultaneous widely-separated showers resulting from the solar-photodisintegration of UHECR nuclei~\cite{GZ}, which was considered by The Large Area Air Shower (LAAS) observatory~\cite{Ochi:1999} but has not been observed. Besides the GZ effect, a large class of theorized physical mechanisms exist in literature that motivate wavefronts of UHECR particles and nuclei arriving at Earth simultaneously.
A comprehensive list includes relativistic dust grains~\cite{HeavyCR, Dust, SuprathermalGrains, InterstellarGrains}; 
cosmic electromagnetic cascades from ultra-high energy gamma-ray pair-production (\emph{pre-showers})~\cite{GammaAbsorption, GammaAbsorption2, GammaAbsorption3, GammaOpacity, SecondaryProduction};
super-GZK neutrinos (the \emph{Z-Burst scenario})~\cite{SuperGZK, NeutrinoAbsorption, NeutrinoDM, NeutrinoDM2}; 
extra-dimensions and localized gravity (the \emph{Gravi-Burst scenario})~\cite{GraviBurst};
and lastly a collection of so-called \emph{top-down exotics} that result in Standard Model particles through radiation, annihilation or collapse processes~\cite{DomainsStrings, DomainsStrings2, Defects, Strings, Monopolonium, MonopoleAnnihilation, SuperconductingStrings, StringsUHECR, UHECRStrings, Vortons, Vortons2, Necklace, BlackHoleNeutrinos, SuperheavyDM}.

To date, no evidence of these mechanisms have been experimentally confirmed, possibly owing to the limitations of existing observational technology.
Many of these mechanisms would predominantly initiate, or be otherwise created through situations far outside the heliosphere.
However, for those that initiate within the confines of the solar system, to wit this set includes the GZ effect, the potential observational rate can be dependent on whether the progenitor particle approaches the Earth along a trajectory that passes near the Sun as opposed to those where the progenitor approaches from the dark side of the planet.
Other mechanisms, such as pre-showers, are most likely to initiate in the extended atmosphere.
In this paper, we illustrate the potential discovery power of a global sensor network by estimating its sensitivity to widely separated simultaneous showers, using those generated by the GZ effect as a benchmark.
We choose to focus on the GZ effect both because it is well anticipated from Standard Model physics, and because it sets a soft lower-limit of detectable global-scale phenomena.
In our Statistical~Analysis~(\S\ref{subsec:StatisticalAnalysis}), we consider the inclusion of such processes that are driven by interactions with solar wind or radiation, or those which likely initiate near Earth as a potential ``booster'' to the observational sensitivity we find for the GZ effect.

\section{Methods}
\label{sec:Methods}

Estimating the sensitivity of a global network to the GZ effect requires an understanding of the combinatoric background from randomly coincident showers~(\S\ref{subsec:CombinatorialBackground}),  a calculation of the rate at which the GZ effect produces dual cosmic rays incident on the Earth~(\S\ref{subsec:GerasimovaZatsepinEffect}), a description of the expected signature of extended air showers,~(\S\ref{subsec:ExtensiveAirShowerParameterization}), and a model of the expected distribution of sensors in various global network configurations, ranging from realistic to optimistic~(\S\ref{subsec:CRAYFISSensitivity}).

\subsection{Combinatorial Background}
\label{subsec:CombinatorialBackground}

The global rate of cosmic ray shower events is quite substantial~\cite{PDG}, and the chances of detecting physically-unrelated but nonetheless simultaneous showers increases with observatory size regardless of the technology employed.  Two features of truly simultaneous showers which will help distinguish them from the combinatoric background are their opening angle $\Delta \psi$, which is expected to be small for nearly parallel correlated showers, and the great-circle separation between the two geographic locations, $\Delta s$, which tends to be larger for unrelated showers.

We assume that the incident flux of cosmic rays is isotropic, though some energy-dependent  anisotropy in the UHECR flux has been identified with higher-order features~\cite{Auger:2014a, Auger:2014b, Auger:2017, Auger:2018, TA:2014, TA:2018, diMatteo:2017}. Unrelated coincident showers are modeled by Monte Carlo methods, generating two random geographic locations drawn with uniform spherical density  with random headings. The opening angle, $\Delta \psi$ is then computed from the difference in headings, and the great-circle separation between the two geographic locations, $\Delta s$, is computed from the well-known haversine distance formula.


\subsection{Gerasimova-Zatsepin Effect}
\label{subsec:GerasimovaZatsepinEffect}

The GZ effect is a potential mechanism for the generation of nearly-simultaneous, but greatly separated showers.  UHECR nuclei are split by solar photons on their way to Earth, and the two daughter nuclei are separated by the solar magnetic field.  Below, we describe the probability for photodisintegration, our model of propogation of the daughters in the solar magnetic field, and an efficient algorithm to generate simulated GZ dual showers. Knowledge of the heliospheric magnetic field (HMF) is the dominant source of systematic uncertainty in the entire analysis.

\begin{figure}[h]
\includegraphics[width=0.85\linewidth]{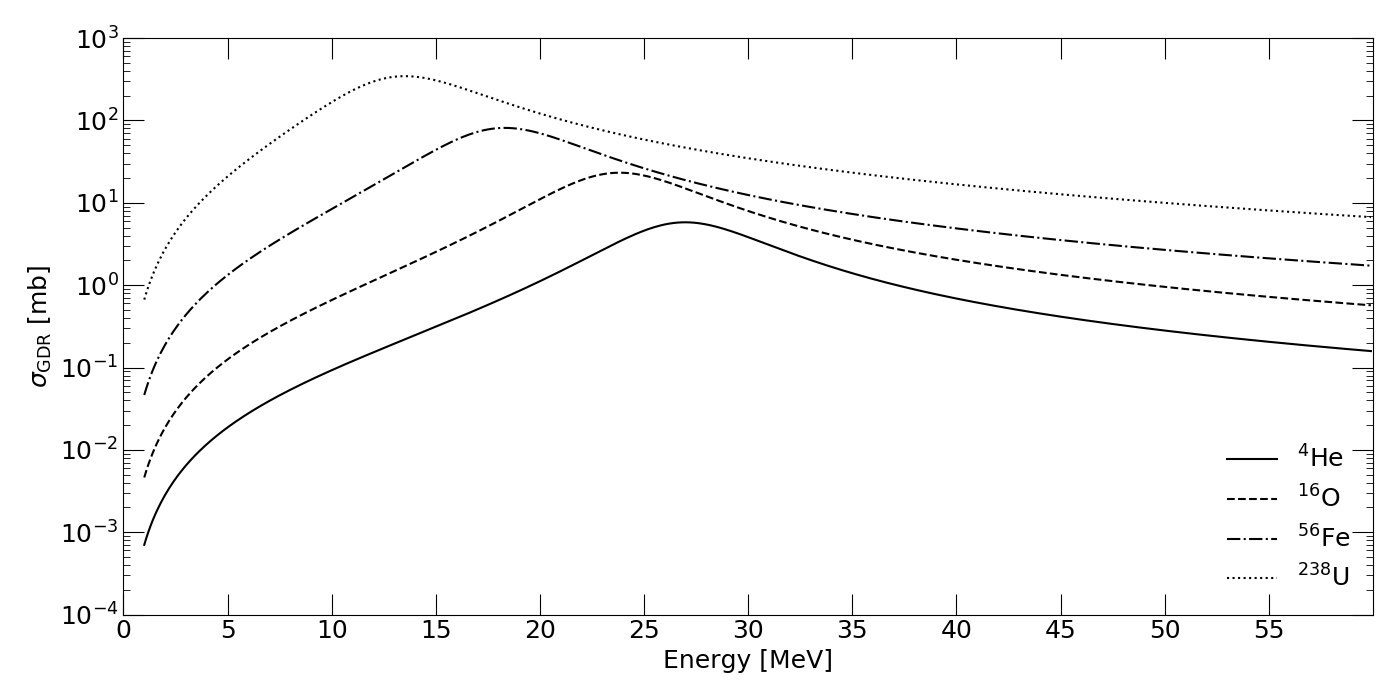}
\caption{        Cross section model for various elements to interact with photons via the Giant Dipole Resonance (GDR).}
\label{fig:xs_gdr}
\end{figure} 

\subsubsection{Photodisintegration}
\label{subsubsec:Photodisintegration}

Following the work of Refs.~\cite{Epele:1998, MedinaTanco:1998}, the mean free path, $\lambda$, for the photodisintegration of a CR nucleus in a photon field density $\mathrm{d}n/\mathrm{d}\epsilon$ with single-nucleon ejection cross section is, 
\begin{equation}\label{eq:Mean_Free_Path}
    \lambda^{-1}(r, \alpha) = \int_0^\infty \mathrm{d} \epsilon \; \frac{\mathrm{d} n(r, \epsilon)}{\mathrm{d}\epsilon} \; \sigma_\mathrm{GDR}\left( \gamma \, \epsilon \, g(\alpha) \right) g(\alpha)
\end{equation}
where $\epsilon$ is the photon energy in the solar rest frame, $r$ is the distance of the nucleus from the Sun, $\sigma_\mathrm{GDR}$ is the photon-nucleus interaction cross section (see Fig.~\ref{fig:xs_gdr}) due to the Giant Dipole Resonance (GDR), $g(\alpha) = \left(1 + \beta \; \mathrm{cos} \; \alpha \right) \simeq 2\, \mathrm{cos}^2 \alpha/2$ is the geometrical Doppler shift of the photon as seen by the nucleus, angle $\alpha$ is the (mis-)alignment between outgoing photon and incoming nuclei momenta, and $\beta$ and $\gamma$ take their usual relativistic definitions. 

To perform numerical simulations, in which the incoming target nucleus is stepped a small, finite distance $\Delta s$ over which $\lambda$ is sufficiently constant, we must then determine the probability for one or more photodissintegration interactions to occur:
\begin{equation}\label{eq:Probability_Step}
    P_{\geq 1}\left(\vec{r} \rightarrow \vec{r}+\Delta s\;\mathrm{d}\hat{r}\right) = 1 - e^{-\Delta s / \lambda(r, \alpha)} \simeq \Delta s / \lambda
\end{equation} 

\subsubsection{Solar System Propogation}
\label{subsubsec:Dynamics}
Gravitational attraction has negligible influence on UHECR propagation over solar system distance scales, as such, the dynamics are governed exclusively by the relativistic Lorentz-force law:
\begin{equation}\label{eq:Lorentz_Force}
    \frac{\mathrm{d} \boldsymbol{p}}{\mathrm{d}t} = q\;\left(\boldsymbol{v} \times \boldsymbol{B}\right),
\end{equation}
where time is measured in the solar rest frame, $\boldsymbol{p} = \gamma m \boldsymbol{v}$, $m$ and $q$ are the mass and charge of the nucleus respectively, and $\boldsymbol{B}$ is the HMF. 
For nuclei in the EeV to ZeV energy range of interest, $10^6<\gamma<10^{10}$, and $.9999999999994\bar{9}<\beta<0.\bar{9}$.
Therefore, with negligible error, $\boldsymbol{\beta} = \hat{\boldsymbol{\beta}}$, and it is possible to re-write Eq.~ \ref{eq:Lorentz_Force} in a numerically-convenient form:
\begin{equation}\label{eq:Custom_Lorentz}
    \frac{\mathrm{d} \hat{\boldsymbol{\beta}}}{\mathrm{d}s} = \frac{Z}{E_{eV}}\left(\hat{\boldsymbol{\beta}} \times c \boldsymbol{B} \right)
\end{equation}
where $s$ is the space-coordinate along the path of the nuclei, $Z$ is the atomic number, $E_{eV}$ the energy of the nucleus in electron-volts and $c$ the speed of light.
The HMF for distances up to 20 AU (approximately the orbit of Uranus) can be modeled as a sum of four primary components~\cite{HMF}. An interpolated map of the HMF sampled every 0.02 AU within 0.6 AU of the Sun and 0.2 AU everywhere else was found to be sufficient for interpolating the HMF to at least 5 figures of accuracy for distances $>$ 0.01 AU of the Sun. For trajectories that fall closer than 0.01 AU, or further than 6 AU, the numerical result is calculated from the full four-component model as needed.

Great care must be taken to ensure numerical accuracy of the propogation.
Spatial precision on the order of meters after propagating 10~AU requires numerical precision at least to the 13th decimal place.  
To establish the accuracy of simulation results, we propogate a relativistic particle under a constant field at its gyroradius~\cite{PDG}:
\begin{equation}\label{eq:Gyroradius}
    r = \frac{1}{c} \frac{ E_{eV} \; \beta }{Z \; B}
\end{equation}
looping it multiple times and comparing the predicted instantaneous radius to the analytical formula. Roughly 100 constant-$B$ cases were validated against Eq.~\ref{eq:Gyroradius} covering the range of HMF magnitudes anticipated from within 2 solar radii of the Sun to 10 AU (the limit of the HMF model applicability), for extreme combinations of $Z$ and $E$.  
Several numerical algorithms were tried, and a Runge-Kutta order 8(5,3) method ``DOP853''~\cite{DOP853}, was found to exhibit the optimal trade-off between performance and accuracy, acruing an error of less than one part in $10^{14}$ after ten revolutions.

During propogation of GZ daughter candidates, the numerical step size is computed at each step to be 1/100$^{\mathrm{th}}$ the analytical gyroradius given in Eq.~\ref{eq:Gyroradius} for the instantaneous HMF strength at the current location, up to maximally 0.01~AU.
As nuclei and ejected nucleons come within one step size of the Earth, the step size is further reduced to fractions of Earth-radii to assure accuracy in geographic termination point.

\subsubsection{Dual Shower Generation Algorithm}
\label{subsubsec:Algorithm} 

A simulation of incoming UHECRs undergoing the GZ effect and resulting in dual showers on the Earth's surface should in principle begin with all possible incoming particles.  However, as the radius of the Earth is roughly $10^{-6}$ the distance of the orbital radius of Neptune, the chance of a cosmic ray propagating over solar system distances to strike the Earth is extraordinarily low. This vanishing efficiency makes such a simulation impractical.

\begin{figure}[h!]
    \centering
    \includegraphics[width=.8\linewidth]{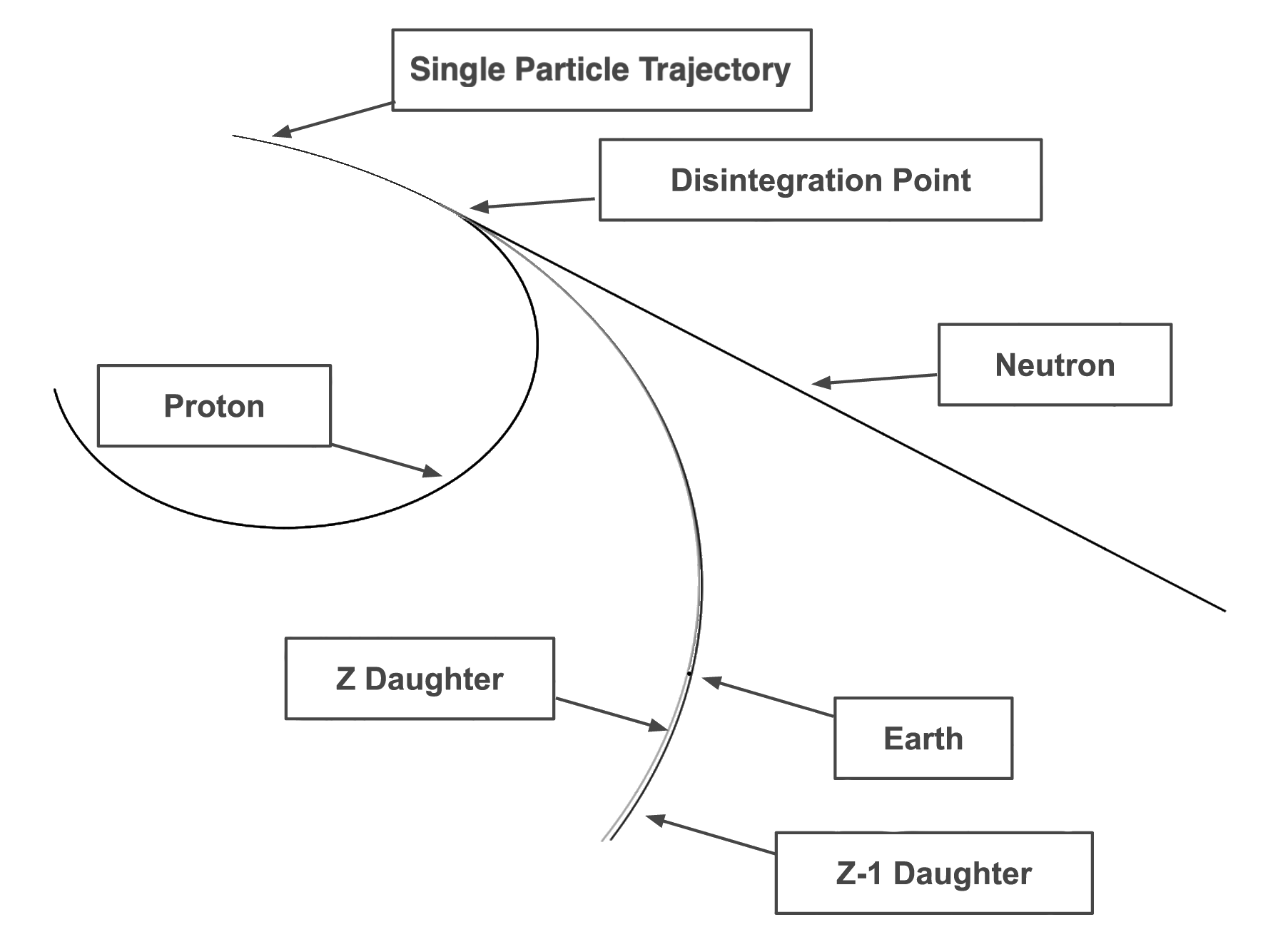}
    \caption{ Simulation of the propogation of a UHECR nucleus before photodisintegration, and the resulting daughter particles. Both scenarios, $Z$ daughter and neutron or $Z-1$ daughter and proton, are shown, though only one would occur.   The magnitude of the curvature is exaggerated for visibility.}
    \label{fig:propagation}
\end{figure}

Instead, we begin with single-particle trajectories which would strike the Earth if not split by interactions with solar photons.  We generate a photodisintegration point by sampling from the probability along each Earth-striking trajectory, and propogate the daughter particles, see Fig.~\ref{fig:propagation}.

Single-particle trajectories were generated from 100,000 geographic seed points selected at random on the surface of the Earth with uniform spherical density. 
The azimuth and zenith angle headings for each point was in turn randomly assigned with uniform hemispherical density away from the local surface. The trajectory is calculated by propogating the single-particle outwards.
For each parent nucleus ($^4$He, $^{16}$O, $^{56}$Fe and $^{238}$U), and for each parent energy ($10^{15}$, $10^{16}$, $10^{17}$, $10^{18}$, $10^{19}$ and $10^{20}$ eV), a negative $Z$ charge is temporally adopted so that the outgoing propagation telemetry would be representative of the incoming positive $Z$ trajectory.

For each discrete step in the outgoing simulation, the probability to photodisintegrate (Eq.~\ref{eq:Probability_Step}) is sampled and stored.
After propagating to 10~AU from the Sun (the limit of the HMF model applicability), the total probability to photodisintegrate along this incoming trajectory is found from summing the stored probabilities.
Because of the inverse square law dependence of the solar photon density, integrating out further (\eg, to Pluto at 40~AU) under zero magnetic field influence results in less than a 1\% increase in total probability, and does not differ from the results of the truncated simulation.

\begin{figure}[h!]
    \centering
    \includegraphics[width=.8\linewidth]{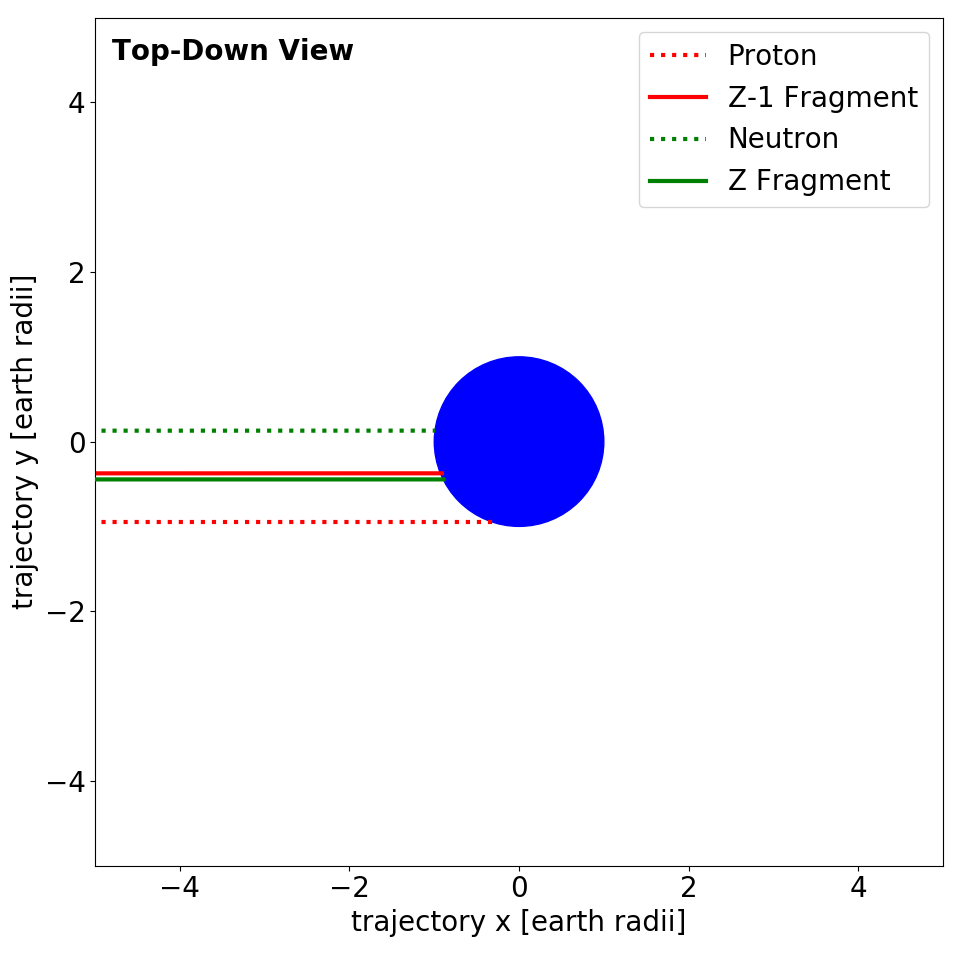}
    \caption{
        An example simulation resulting in dual-showers for both the `p' (red) and `n' (green) channels, [dotted] nucleons, and [solid] nuclear fragments.  
        Channel pairs can either both miss the Earth, single shower (one partner misses), or dual shower. 
        Dual showers can potentially be separated by an entire Earth diameter.  
        The trajectory-$x$ direction is taken as the direction of propagation of the proton, with the $z$ direction parallel to the solar system $z$. 
    }
    \label{fig:dual_showers}
\end{figure}

A random point along this 10~AU seed trajectory (weighted by the stored distribution of step probabilities) is then drawn to serve as the photodisintegration point.
The incoming simulation then begins with the parent nucleus situated at this disintegration point and the simulation splits into two separate outcome channels we denote `p' (for proton and $Z$-1 daughter), and `n' (for neutron and $Z$ daughter).
We assume~\cite{Epele:1998} the parent energy is split between fragments in proportion to nucleon number,
\begin{equation}\label{eq:E_Split}
    \begin{split}
        E_{\mathrm{nucleon}}  &= \frac{1}{A} E_{\mathrm{primary}} \\
        E_{\mathrm{fragment}} &= \frac{A - 1}{A} E_{\mathrm{primary}}.
    \end{split}
\end{equation}
Kinematically, the interaction of an UHECR with a low energy photon ($E_{\gamma} / E_{\mathrm{primary}}~\ll~10^{-12}$) results in negligible transverse momentum, and the resulting nuclear fragments separate within a boundary cone of virtually zero to great accuracy on solar system scales.
Therefore, the daughter fragments take on the initial heading of the parent, and their separation is dominated completely by the HMF.
Each of the nucleon-daughter fragment channels is separately propagated forward, and the simulation ends for a fragment when it has either struck the Earth or passed it, see Fig.~\ref{fig:dual_showers}.

\subsection{Extensive Air Showers}
\label{subsec:ExtensiveAirShowerParameterization}

UHECRs striking the Earth's atmosphere produce extensive air showers, which can be modeled with the the CORSIKA~\cite{CORSIKA} software package. As the computational time for a shower simulation scales roughly linearly with primary energy, running high-statistic simulations for \emph{ad~hoc} primary projectile, incident angle and observational altitude configurations demand significant computational costs.  The statistical analysis of a search for rare events among a high-rate background demands a large number of sample showers.  To ensure computational tractability, a parameterizated particle density as a function of the distance from the shower core is typically fit to samples of showers simulated with CORSIKA  at various energies and for several species. 

We compiled CORSIKA in 64-bit mode with GHEI\-SHA 2002d for a horizontal flat array with thinning (including LPM) support.
As the hadronic interaction model is the dominant uncertainty in Monte Carlo shower codes, we performed each simulation configuration 5 times, one for each available hadronic model:  DPMJET-III (2017.1) with PHOJET 1.20.0, QGSJET 01C (enlarged commons), QGSJETII-04, SIBYLL 2.3c and VENUS 4.12.
The charmed particle / tau lepton PYTHIA option was activated for DPMJET, QGSJET and SIBYLL.
For all non-vertical simulations, the curved atmosphere selection was enabled. Values of essential CORSIKA parameters believed to optimize the realism of the shower simulations with acceptable computational time is provided in Tab.~\ref{tab:corsika}

For each of the 5 hadronic interaction models, 100 shower simulations were run for each of 5 shower-primary projectiles (photon, proton, Helium, Oxygen and Iron), for each of 8 shower-primary energies ($10^{14}$, $10^{15}$, $10^{16}$, $10^{17}$, $10^{18}$, $10^{19}$, $10^{20}$ and $10^{21}$ eV), and for each of 4 incident zenith angles (vertical, 30$^{\circ}$, 60$^{\circ}$ and 80$^{\circ}$).
For vertical simulations, 10 observation altitudes were specified at 0, 0.5, 1, 1.4, 2, 5, 10 and 20 km a.s.l.
For angled-incident simulations where the curved atmosphere option limits observation altitudes to 1 per simulation, 5 simulations were performed for altitudes 0, 1, 2, 5 and 10 km (instead of all 5 hadronic models, only one was chosen at random per simulation).
In total, roughly 100,000 simulations were performed.

\begin{figure}[h!]
  \includegraphics[width=.8\linewidth]{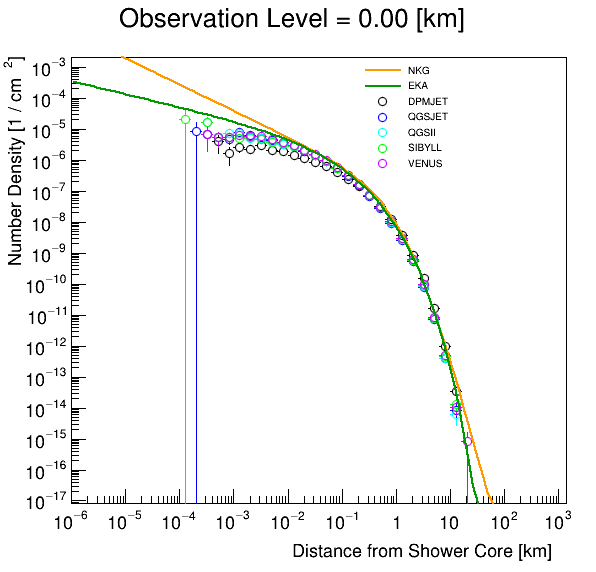}
  \includegraphics[width=.8\linewidth]{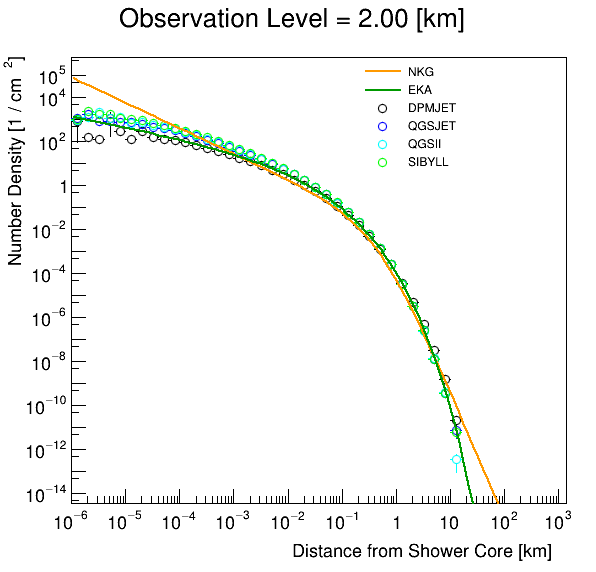}
    \caption{
        Lateral density models for muons from $10^{14}$ eV Helium (top) and photons from $10^{17}$ eV Oxygen (bottom) compared with CORSIKA simulations using various high-energy hadronic interaction packages and the settings listed in Table~\ref{tab:corsika}.  In general, the NKG (orange) function is found to over-predict densities near the shower core and taper off slower than simulation results predict.  We find better simulation agreement with the proposed alternate EKA model in green (Eq.~\ref{eq:EKA}).}
    \label{fig:Model_Test}
\end{figure}

While several standard paramaterization exist, it is not uncommon for both experiments and Monte Carlo simulations to find deviations from the analytically-motivated NKG~\cite{NKG} distribution near and far from the core~\cite{AugerLDF, Hillas:1977, Fenyves:1988}.   To accommodate these deviations, it is typical to develop variants of these parameterizations~\cite{GaisserHillas, Schiel:2006}.   Here, we develop a novel longitudinal shower density particle parameterization, which we name EKA, for photons and muons as a function of primary atomic mass, energy, incidence angle and observational altitude that describes well both the shower core and extended tails:
\begin{equation}\label{eq:EKA}
    \rho(r; a_{n}) = e^{a_0} r^{-a_1} exp\left(-\frac{r^{a_3}}{e^{a_2}}\right) 
\end{equation}
where the four $a_{n}$ coefficients are best-fit as a function of first-order factors of transformed primary mass number, $A$, primary energy, $\epsilon$, and altitude of observation, $h$ as:
\begin{align}\label{eq:Expansion}
    \begin{split}
        a_{n}(A^*, \epsilon^*, h^*) \cong \; & c_n^0 + \\
                                & c_n^1 A^*   + c_n^2 \epsilon^*   + c_n^3 h^*   + \\
                                & c_n^4 A^*\epsilon^* + c_n^5 A^*h^* + c_n^6 \epsilon^* h^* + \\
                                & c_n^7 A^*\epsilon^*h^*
    \end{split}
\end{align}
where,
\begin{align}\label{eq:Coeff_Transform}
    \begin{split}
        A^* &= \mathrm{ln}\;(A + 1) \\
        \epsilon^* &= \mathrm{log_{10}}\;(\epsilon / 10^{18}) \\
        h^* &= 100\;\left(1 - \mathrm{log_{10}}\;(10 - h_\mathrm{eff}/10) \right)
    \end{split}
\end{align}
with energy in electron-volts, altitude in kilometers and for vertical showers, $h_\mathrm{eff}$ is simply the altitude of observation, $h$.
In general, for showers inclined by an angle $\theta_{0}$ above the observation horizon at altitude $h$, and coming from azimuthal direction $\phi_{0}$, 
\begin{equation}\label{eq:Slant_Transform}
    h_\mathrm{eff} = h' + r\;\mathrm{sin}\;\theta_{0}\;\mathrm{cos}\;(\phi - \phi_{0})
\end{equation}
where $h'$ in turn is a function of the altitude of the first interaction, $h_{0}$:
\begin{equation}\label{eq:H_Prime}
    h' = h_0 - (R_E + h) \left( \sqrt{\left(\frac{R_E + h_0}{R_E + h}\right)^2 - \mathrm{cos}^2\;\theta_0} - \mathrm{sin}\;\theta_0 \right)
\end{equation}
where $R_E$ is the radius of the Earth in kilometers, and the average first interaction altitude, $h_0$, is modeled from simulation to good agreement as:
\begin{equation}\label{eq:H_0}
     h_0(A^*, \epsilon^*) \cong d_0 + d_1 A^* + d_2 \epsilon^* + d_3 A^* \epsilon^*
\end{equation}
with coefficients $d_n$ provided in Tab.~\ref{tab:coeff}.  Example fits of our EKA model and the NKG model to CORSIKA samples are shown in Fig.~\ref{fig:Model_Test}. The EKA model is used to describe the expected particle densities in all subsequent simulations.

\begin{table}
\caption{Best-fit coefficients to Eq.\ref{eq:H_0}}
\label{tab:coeff}
\begin{tabular}{llll}
\hline\hline
$d_0$ \ \ & $d_1$\ \  & $d_2$ \ \ & $d_3$ \\
\hline
24.1 & 2.74 & 0.550 & -0.061 \\
\hline\hline
\end{tabular}
\end{table}

\subsection{CRAYFIS Sensitivity}
\label{subsec:CRAYFISSensitivity}

In this subsection, we describe our modeling of the effective observing power of a globally distributed array, and our methods for simulating dual showers impacting on the array, for two CRAYFIS array size and density scenarios.

\subsubsection{Effective Area}

The crucial metric for a cosmic ray observatory is the  \emph{effective area}, $A\epsilon$, the product of the total observing area and the efficiency to detect a shower.  The effective area depends on the efficiency of individual sensors, as well as their density and location.

\begin{figure}[h!]
    \centering
    \includegraphics[width=\linewidth]{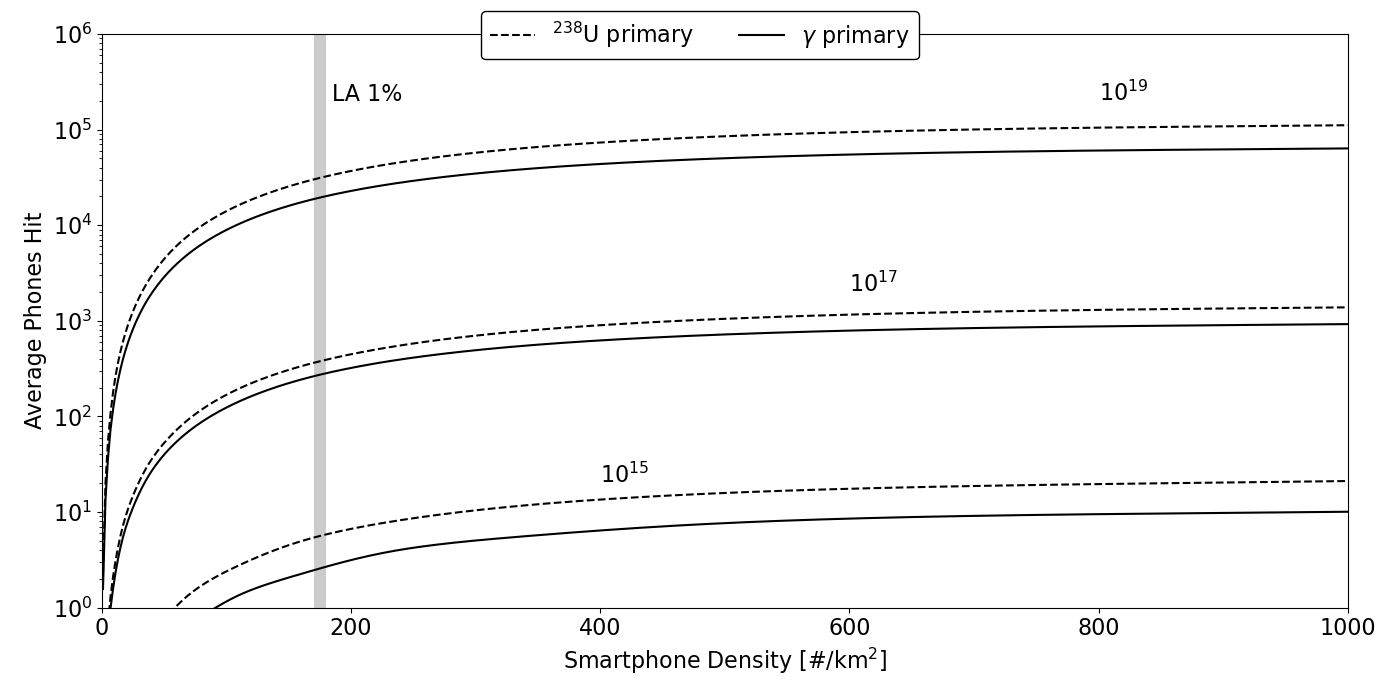}
    \includegraphics[width=\linewidth]{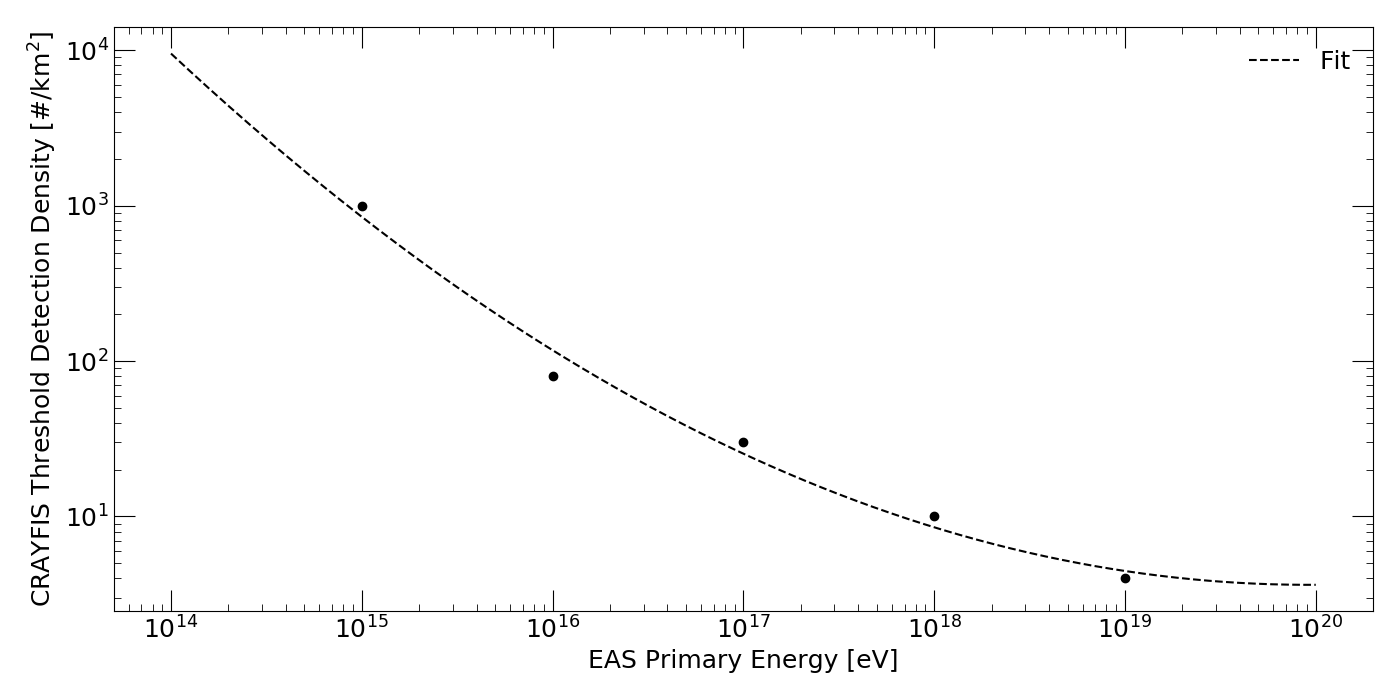}

    \caption{
        Top, the average number of smartphones registering a ``hit'' from photons and muons as a function of smartphone density (linear scale) within 10 km radius of an EAS.
        Primary energies are listed by each curve.
        %
        %
        The vertical bar denotes the smartphone density of 1\% user adoption in the city of Los Angeles.  Bottom, the threshold smartphone density versus primary energy, $\rho_{\textrm{thresh}}(E)$, fit to a quadratic function.
    }
    \label{fig:lab_sensitivity}
\end{figure}

In our previous work~\cite{crayfis}, we estimated the effective area of individual smartphone camera sensors to photons and muons using measurements of cosmic rays, radioactive sources and beam tests. The effective area for photons from laboratory sources incident normal to the camera sensor was found to typically range from $A\epsilon \sim 10^{-5}$~to~$10^{-4}$~cm$^2$, and accelerator-based tests place a conservative effective area of ${\sim}0.05$ cm$^2$ for muons.  To identify a shower, Ref~\cite{crayfis} proposed that a threshold of five geographically clustered smartphone hits within a 100~ms window was sufficient to reduce the background from combinatorics due to the overwhelming rate of cosmic muons from low-energy showers.    As the number of particles in a shower scales with the energy of the primary cosmic ray, the efficiency for a cluster of phones to detect a shower will depend on both the energy of the shower and the density of the smartphones.  We use Monte Carlo simulations to identify the thresold density above which the a cluster of smartphones are essentially fully efficient, $\rho_{\textrm{thresh}}(E)$, as a function of the primary cosmic ray energy, see Fig.~\ref{fig:lab_sensitivity}.

The effective area of the entire array is very sensitive to the number of localized patches of users that are above the threshold density.   As the number of users who may lend their device to the network is difficult to anticipate, we explore two scenarios for the average CRAYFIS smartphone density $\langle \rho_C \rangle_i$ which is factored,

\begin{equation}
    \begin{split}
    \langle \rho_C \rangle_i &= \frac{N_\mathrm{people}}{A_i} \left\langle \frac{N_\mathrm{smartphones}}{N_\mathrm{people}} \right\rangle \left\langle \frac{N_\mathrm{CRAYFIS}}{N_\mathrm{smartphones}} \right\rangle \\
                                          &= \rho_i \langle \xi \rangle
    \end{split}
\end{equation}
where $A_i$ is land area so that $\rho_i$ is the population density, and $\langle \xi \rangle$ is the product of scenario-dependent quantities:
\begin{itemize}
    \item `Scenario U' (the \underline{U}pper-limit scenario):
    \begin{itemize}[label={}]
        \item $\left\langle \frac{N_\mathrm{smartphones}}{N_\mathrm{people}} \right\rangle_\mathrm{U} = 1$
        \item $\left\langle \frac{N_\mathrm{CRAYFIS}}{N_\mathrm{smartphones}} \right\rangle_\mathrm{U} = 1$
        \item $\implies \langle \xi \rangle_\mathrm{U} = 1$
    \end{itemize}
    
    \item `Scenario P' (the \underline{P}ragmatic scenario):
    \begin{itemize}[label={}]
        \item $\left\langle \frac{N_\mathrm{smartphones}}{N_\mathrm{people}} \right\rangle_\mathrm{P} \simeq \frac{2.87\times10^{9}}{7.79\times10^{9}} = 0.368$
        \item $\left\langle \frac{N_\mathrm{CRAYFIS}}{N_\mathrm{smartphones}} \right\rangle_\mathrm{P} \simeq \frac{1}{1,000}$
        \item $\implies \langle \xi \rangle_\mathrm{P} \simeq 3\times10^{-4}$
    \end{itemize}
\end{itemize}

The maximal sensitivity of CRAYFIS is explored in `Scenario U', whereas a more realistic potential future scenario is explored in `Scenario P'.
In the latter case, the number of smartphone users world-wide, $N_\mathrm{smartphones}$, and the total world population, $N_\mathrm{people}$ are estimated for 2020 in Refs.~\cite{quora, worldometers}, and the fraction of smartphone users with the CRAYFIS app is taken so that the total number of CRAYFIS devices is $\mathcal{O}\left[10^{6}\right]$, a tiny fraction of the billions of active devices.

The effective area contribution from a patch of land is also dependent on the fraction of 24 hours where users are taking data at the same time, $\langle D \rangle = \langle T_\mathrm{data} / \mathrm{24\;hr} \rangle$, where $\langle D \rangle_\mathrm{U} = 1$ and $\langle D \rangle_\mathrm{P} \simeq 6 / 24 = 0.25$ are taken for `Scenario U' and `Scenario P' respectively.

\begin{figure}[h!]
    \centering
    \includegraphics[width=\linewidth]{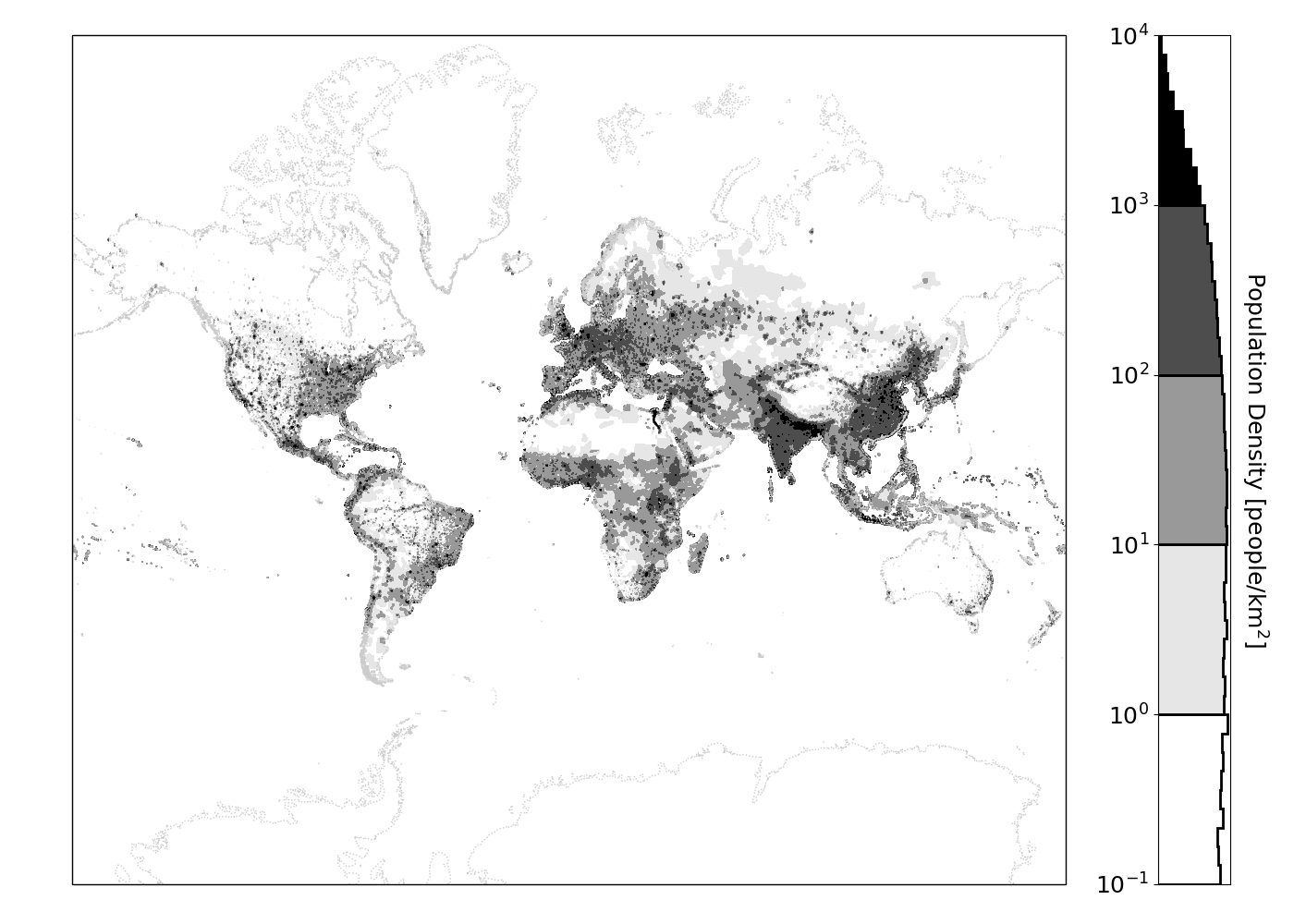}
    \caption{
        The estimated world population density by 2020 in $0.25^{\circ}\times0.25^{\circ}$ bins (dataset courtesy of Ref.~\cite{GPWv4}).
        The horizontal direction of the colorbar indicates the relative frequency of population density bins on a log scale that is not shown.
    }
    \label{fig:population_density}
\end{figure}

The effective areas for both scenarios are then evaluated using an estimated 2020 world-wide population dataset~\cite{GPWv4}, see Fig.~\ref{fig:population_density}.

\begin{figure}[h!]
    \centering
    \includegraphics[width=\linewidth]{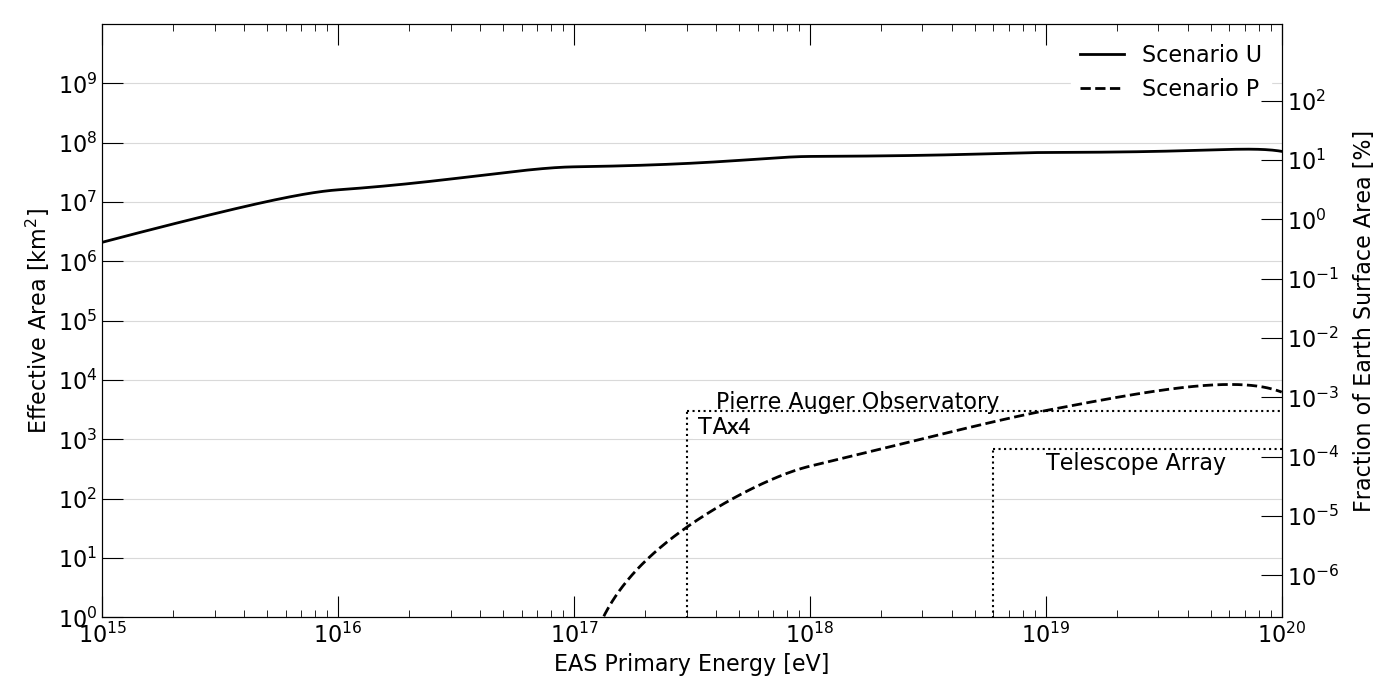}
    \caption{
      The estimated effective area for a CRAYFIS array (`Scenario U', solid), and a ${\sim}$1~million user expectation (`Scenario P', dashed).
        Also shown are the surface detector effective areas of the Pierre Auger and Telescope Array Observatories (dotted).
        AugerPrime is expected to cover nearly the same region on this figure, which is also shared with the reach of the Telescope Array upgrade,``TAx4.''
    }
    \label{fig:Ae_Earth}
\end{figure}

The world-wide network effective area is:
\begin{equation}\label{eq:Ae}
    A\epsilon(E) = \langle D \rangle R_E^2 \Delta\theta \Delta\phi \sum\limits_i \sin\theta_i \mathcal{H}\left(\rho_i \langle \xi \rangle - \rho_\mathrm{thresh}(E) \right)
\end{equation}
where $R_E$ is the radius of the Earth, and ($\Delta\theta$, $\Delta\phi$) are the radian bin-widths for the $i^{\mathrm{th}}$ population density bin, the threshold density $\rho_\mathrm{thresh}(E)$ is extracted as described above, and a Heaviside step-function only considers fully-efficient patches of land.  The resulting effective areas for both scenarios are shown in Fig.~\ref{fig:Ae_Earth} as a function of primary particle energy.

\subsubsection{Dual Shower Simulations}
\label{subsubsec:LandPatchDistribution}

The effective area calculation is most directly useful for estimates of the power of an observatory to see single EAS events.  In order to model dual-EAS events, we must consider the relative distances between the CRAYFIS-instrumented patches of land. 

The GZ effect simulation in \S\ref{subsec:GerasimovaZatsepinEffect} describes the probability of simultaneous dual-showers separated by distance $\Delta s$. The rate of observed dual showers must also account for the odds for a pair of showers to be simultaneously detected, based on the expected global population density distribution.

\begin{figure}[h!]
    \centering        
    \includegraphics[width=\linewidth]{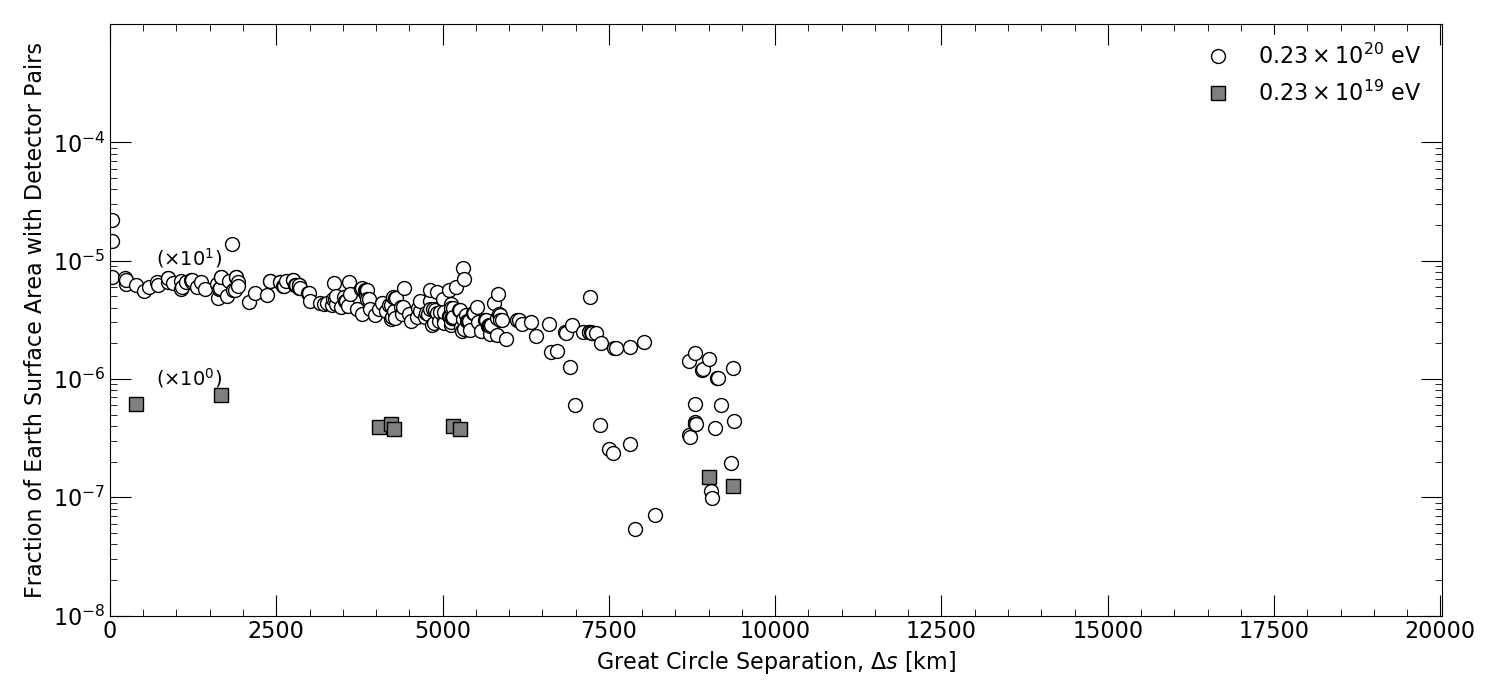}
    \caption{
    The average fraction of Earth's surface area covered by population centers with sufficient CRAYFIS user density, separated by $\Delta s_\mathrm{geo}$ at any given moment under scenario ``P'' for ${\sim}10^{6}$ world-wide nighttime CRAYFIS users averaged over one year.
    Below $0.23\times10^{19}$~eV, no pairs of geographic locations have sufficient population density (according to Ref.~\cite{GPWv4}) to detect dual-EASs at night.
    The maximum separation distance of ${\sim}$20,000~km corresponds to half the circumference of the Earth.
    }
    \label{fig:scenario_sep_area}
\end{figure}

We estimate these odds by computing the fractional surface area  of patches with sufficient CRAYFIS cluster density for detection separated by same distance $\Delta s$ to the total surface area of the Earth. For ``Scenario U,'' where the CRAYFIS array is assumed to be operating at 100\% capacity 24 hours a day, 365 days a year, the land surface area of CRAYFIS cluster-pairings for dual-showers separated by $\Delta s$ is simply the combined surface area of geographic locations separated $\Delta s_\mathrm{geo} = \Delta s$ from each other.
On the other hand, for ``Scenario P,'' where it is assumed that CRAYFIS users are only recording data for 6 hours a day, between 11pm and 5am local time, a numerical simulation is performed which accounts for the day/night availability of the phones as the patches rotate with the Earth, averaged over one year in 15 minute incremenets, see Fig.~\ref{fig:scenario_sep_area}.


\section{Results}
\label{sec:Results}

\subsection{Combinatorical background}
\label{subsec:Background}

\begin{figure}[h!]
    \centering        
    \includegraphics[width=.49\linewidth]{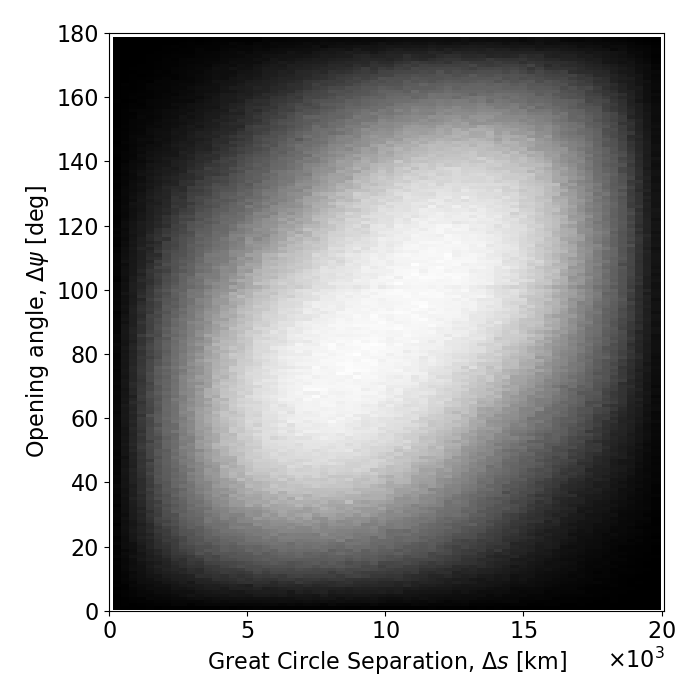}
    \includegraphics[width=.49\linewidth]{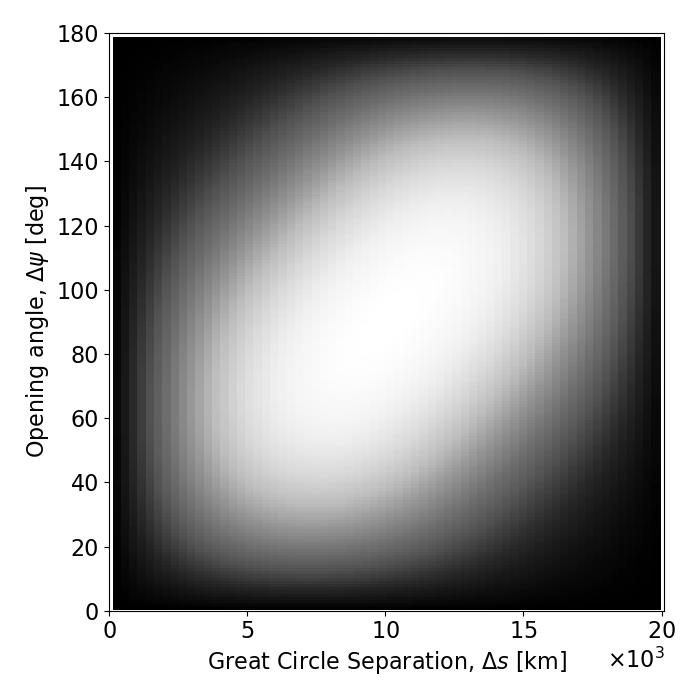}
    \caption{
    The combinatorial background distributions of any two time-coincident cosmic rays as a function of the opening angle, $\Delta \psi$, and the great-circle separation between their geographic locations, $\Delta s$. Left shows the Monte Carlo joint-PDF for both variables; right is the analytical model described in the text.}
\label{fig:bg_model}
\end{figure}

The Monte Carlo model of the probability of randomly coincident cosmic rays is shown in Fig.~\ref{fig:bg_model} as a function of the opening angle, $\Delta \psi$, and the great-circle separation between their geographic locations, $\Delta s$.  An analytical representation which agrees well with Monte Carlo results of $10^7$ simulated cosmic ray pairs is:
\begin{equation}\label{eq:jointPDF}
    \begin{split}
    \mathrm{PDF}(\Delta s, \Delta \psi) = & \frac{1}{4 R_E} \sin\left(\frac{\Delta s}{R_E}\right) \sin\left(\Delta \psi\right) \\
                                                               & \left[ 1 + \frac{3}{4}\cos\left(\frac{\Delta s}{R_E}\right)\cos\left(\Delta\psi\right) \right].
    \end{split}
\end{equation}

Constraints on the opening angle $\Delta\psi$ are the most powerful background discriminator as it is rare for random simultaneous showers to also be near-parallel.  A model of the background  such that $0 < \Delta\psi < \Psi$, for some limit $\Psi$, Eq.~\ref{eq:jointPDF} becomes,
\begin{equation}\label{eq:background}
    \begin{split}
    \mathrm{PDF}(\Delta s; &\Delta\psi \le \Psi) = \\
                                         & \frac{1}{4 R_E}\csc^2\left(\frac{\Psi}{2}\right)\sin\left(\frac{\Delta s}{R_E}\right) \\
                                         & \left[1 - \cos \Psi + \frac{3}{8}\sin^2\Psi \cos\left(\frac{\Delta s}{R_E}\right) \right]
    \end{split}
\end{equation}

\subsection{Expected Rates of GZ Dual Showers}
\label{subsec:GZSimulations}

The methods described earlier are used to estimate the rate at which UHECRs photodisintegrate and produce daughters which both strike the Earth's atmosphere.

\begin{figure}[h!]
\includegraphics[width=0.95\linewidth]{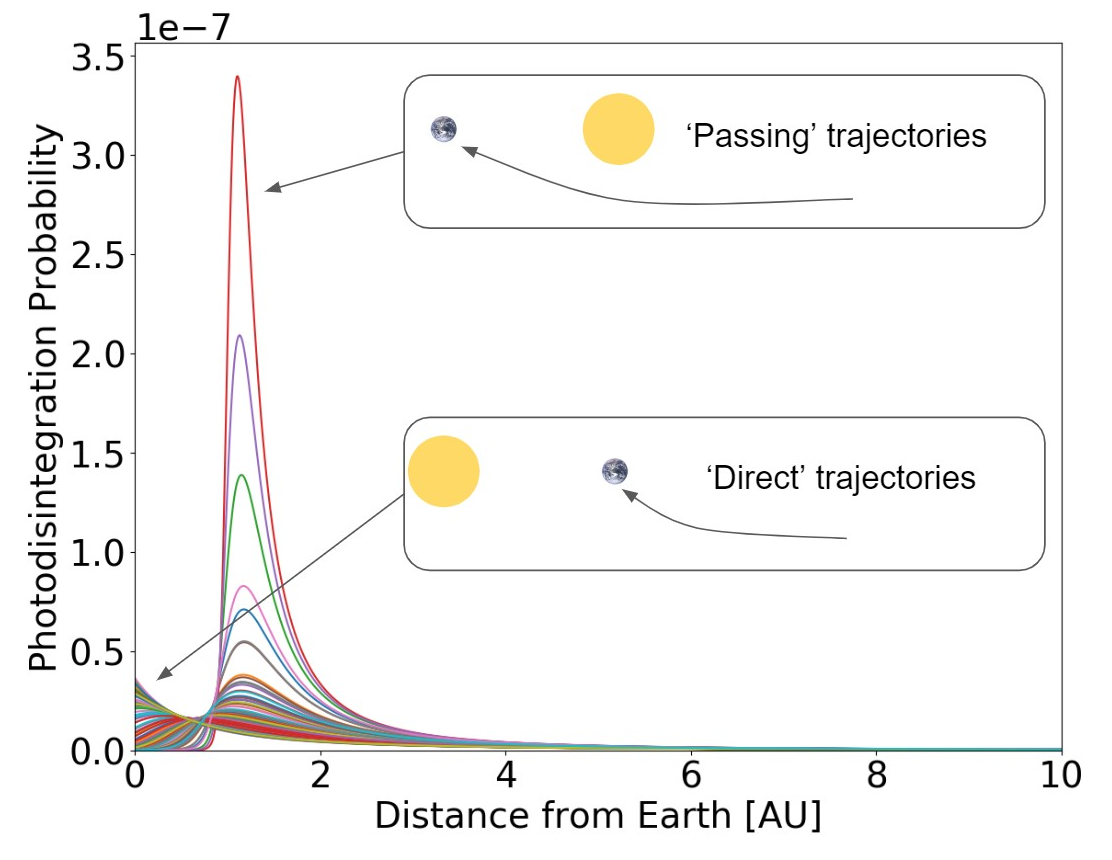}
\caption{        Photodisintegration probability versus distance from Earth for many EeV Oxygen nuclei propagated through the solar system (paths go from right-to-left, where the Earth is at $x=0$, and Sun at $x=1$); different colors are used to distinguish the trajectories.  Trajectories passing near the Sun before heading to Earth (right to left in the Figure) benefit from both head-on incident geometry ($\alpha \sim 0^{\circ}$) and high photon flux, and exhibit the greatest likelihood of disintegration only to fall dramatically upon passing the Sun ($\alpha \sim 180^{\circ}$). In constrast, trajectories for nuclei which arrive at the Earth without passing by the Sun have a probability to disintegrate entirely driven by proximity to the Sun. }
\label{fig:disint}
\end{figure}

\begin{figure}[h!]
    \centering        
    \includegraphics[width=\linewidth]{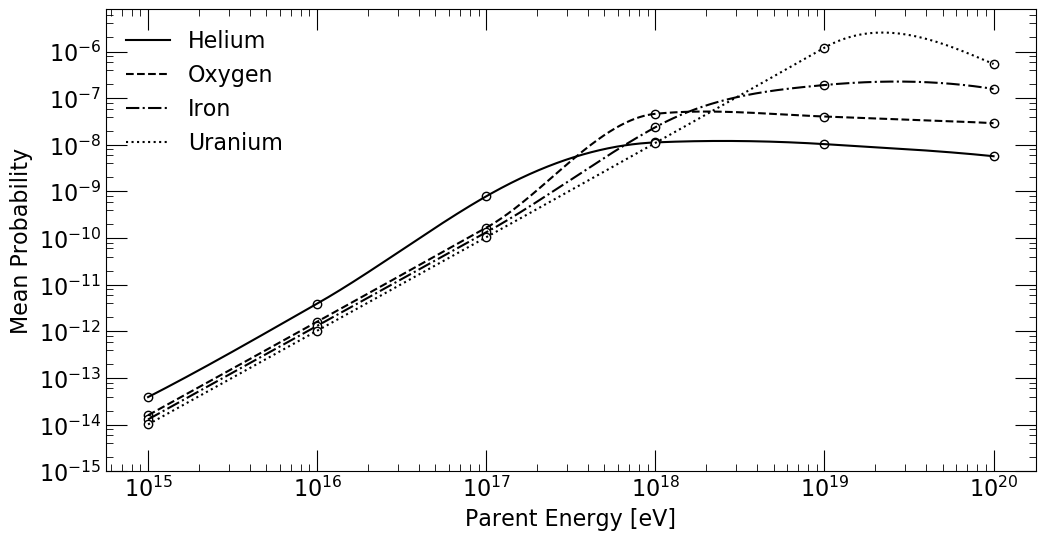}
    \includegraphics[width=\linewidth]{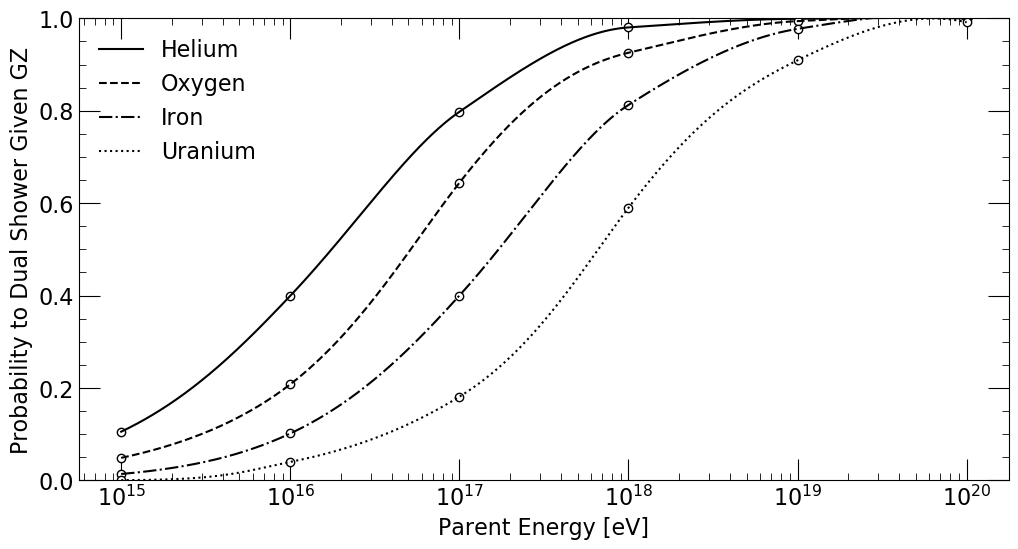}
    \caption{
    Top, the mean probability for any Earth-bound UHECR to photodisintegrate along its trajectory via the Giant Dipole Resonance.
    Bottom, the probability for both daughter products to strike the Earth,  given photodisintegration.
    }
    \label{fig:gz_results1}
\end{figure}

\begin{figure}[h!]
    \centering        
    \includegraphics[width=\linewidth]{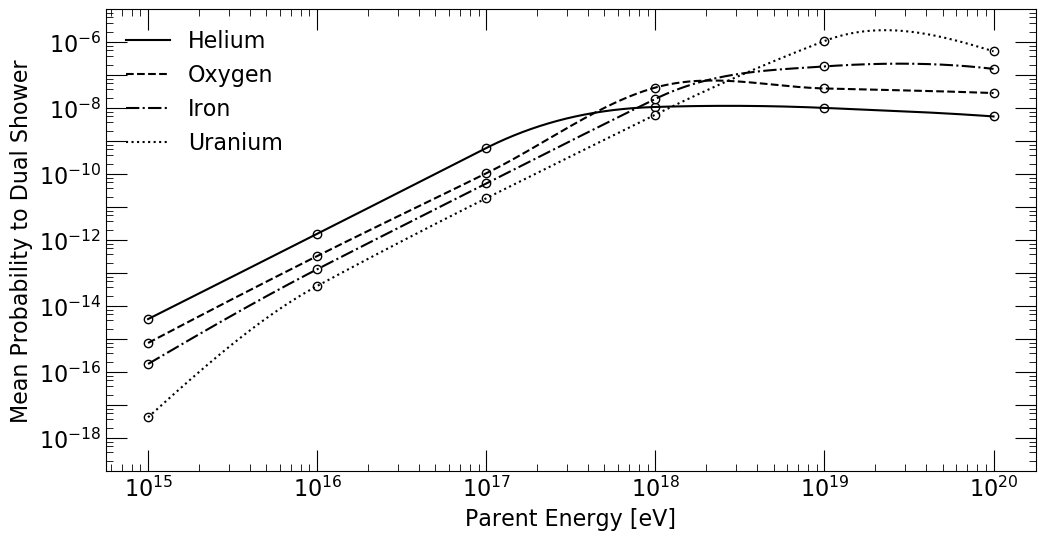}
    \includegraphics[width=\linewidth]{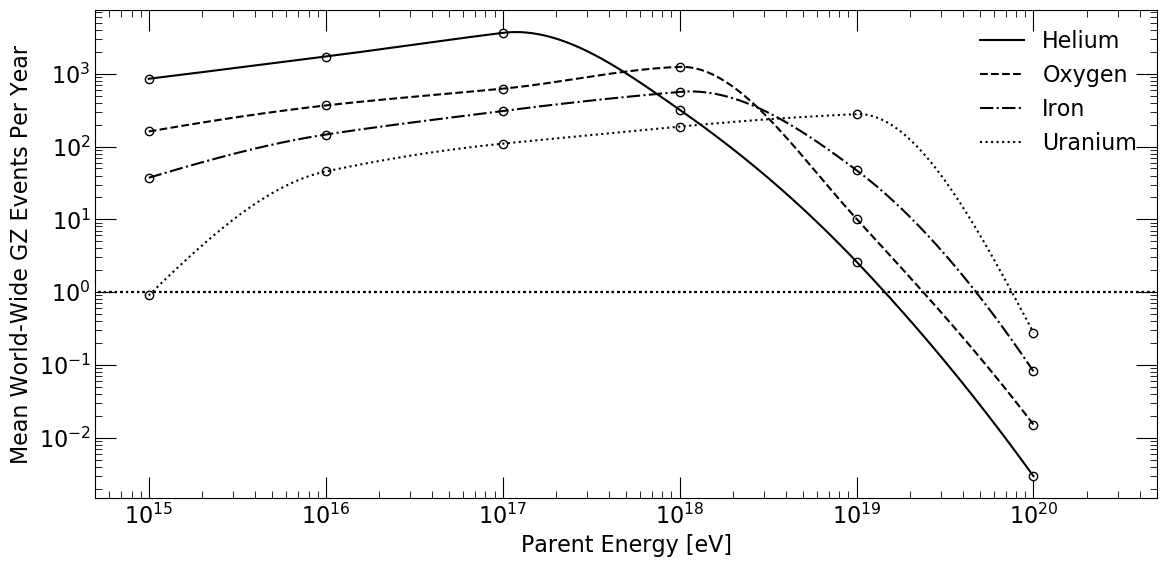}
    \caption{
    Top, the mean expected fraction of Earth-bound cosmic rays to produce dual-EASs, calculated as the product of the probability for UHECRs to photodisintegrate along the trajectory (Fig.~\ref{fig:gz_results1}a) and their probability to dual-EAS given photodisintegration (Fig.~\ref{fig:gz_results1}b).
    Bottom, the product of the top with the  annual cosmic ray flux as estimated in Ref.~\cite{PDG} on the surface of the Earth.
    A horizontal line is drawn to indicate the 1 event per year threshold.
    As the atomic number dependence of UHECRs is poorly known, each element is listed as if they were the only species; therefore, these curves represent the absolute upper-limits for these averages.
    }
    \label{fig:gz_results2}
\end{figure}

As an example, the probability for an Oxygen UHECR nucleus to photodisintegrate is shown in Fig.~\ref{fig:disint} as a function of the distance from Earth.
The average probability over all incoming trajectories to photodisintegrate  and the subsequent fraction of these events that produce daughter particles which both strike the Earth is presented in Fig.~\ref{fig:gz_results1}a.

Photodisintegration via the Giant Dipole Resonance is most effective on nuclei with energy in excess of $10^{18}$~eV, generally plateauing somewhere between a 1~in~10~million likelihood for heavy elements, and 1~in~100~million for light elements.
Somewhat conveniently, the gyroradius for fragments and protons at this energy range become large enough to almost ensure dual-EASs (Fig.~\ref{fig:gz_results1}b)---as the gyro-radius scales inversely with atomic number, $Z$; this is more the case for lighter elements than heavier.

The mean expected fraction of Earth-bound cosmic rays to produce dual-EASs (Fig.~\ref{fig:gz_results2}a) is the product of the fraction of UHECRs to photodisintegrate along their trajectory (Fig.~\ref{fig:gz_results1}a) and the probability to have both daugthers strike the Earth (Fig.~\ref{fig:gz_results1}b).   The small probability for a GZ Effect candidate event is balanced by the very large size of the Earth and the high flux at lower energies, resulting in a large yearly global flux, as shown in Fig.~\ref{fig:gz_results2}b under the assumption that the cosmic rays are of a single elemental nucleus species.

\begin{figure}[h!]
    \centering        
    \includegraphics[width=\linewidth]{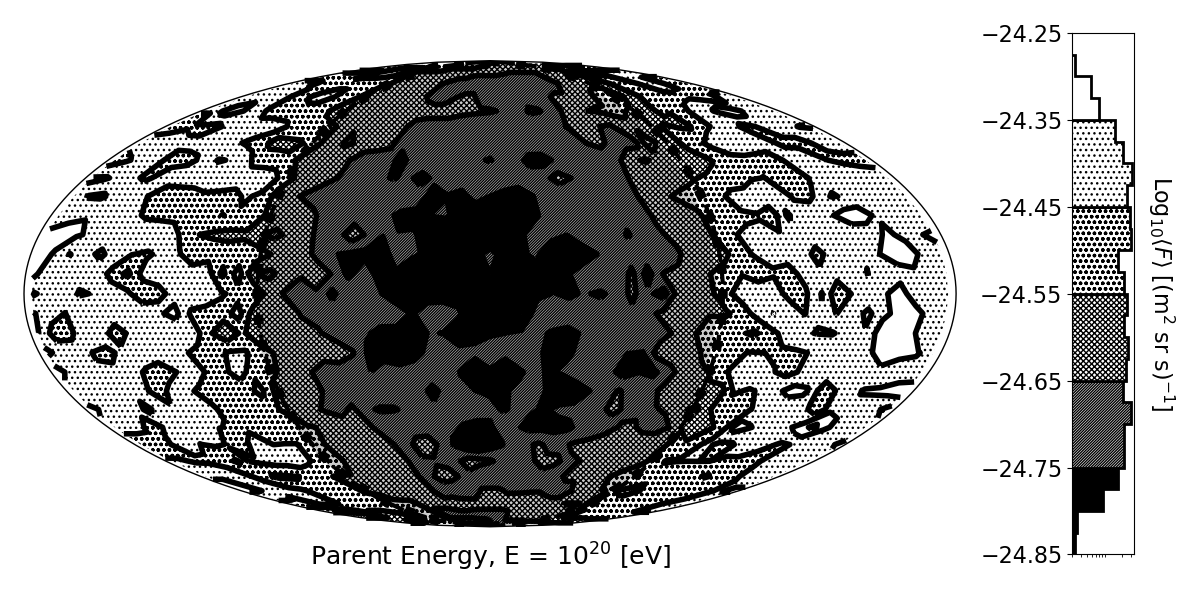}
    \includegraphics[width=\linewidth]{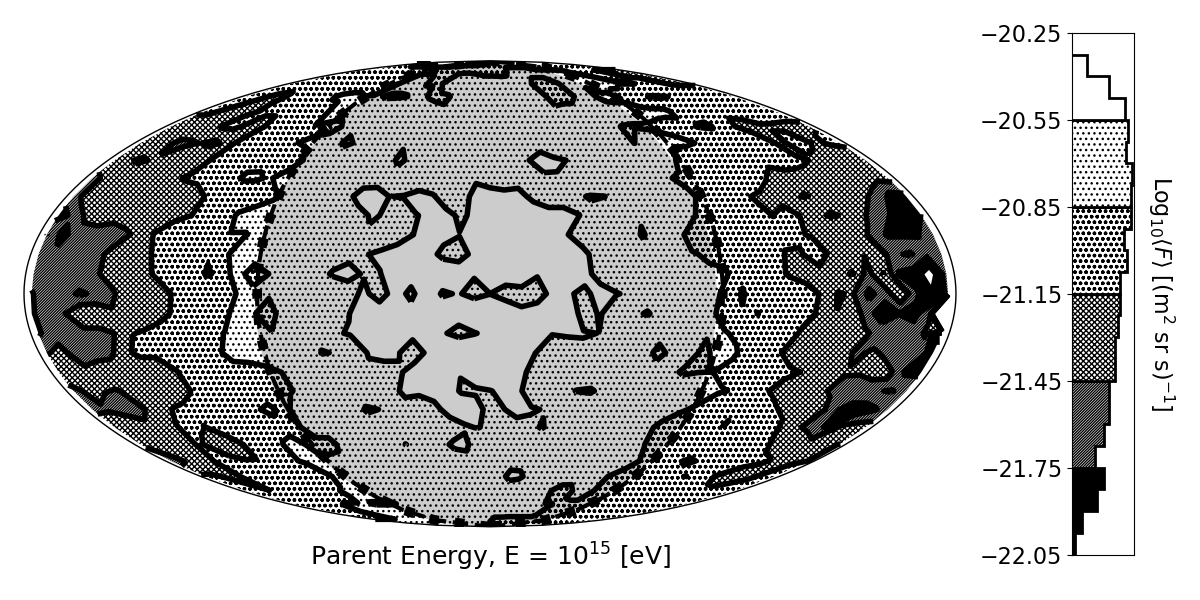}
\caption{
GZ Effect flux on Earth averaged over parent UHECR atomic number for two sample primary energies.
    The central circle identifies the parts of the globe experiencing night (6pm to 6am), \emph{i.e.} when CRAYFIS users are expected to most likely be taking data.
    The equator is aligned with the ecliptic plane with North-South perpendicular to it.
    Eastward rotation is to the right.
    Average flux for $10^{20}$ eV UHECR parent nuclei are shown above, and $10^{19}$ eV is shown below. 
    Color bars indicate the relative land-area coverage on a log scale that is not shown.
    See text for discussion.
    }
    \label{fig:gz_results3a}
\end{figure}

As the GZ probability depends on the atomic number, a calculation of the overall rate of dual showers from GZ effects over the surface of the Earth must fold in the expected relative abundance of various nuclei. However, as this is not precisely known, each parent element's flux contribution was estimated with relative weights that favor lighter elements  (e.g. Oxygen, relative weighting $10^{-1}$) and suppress heavier elements (e.g. Uranium, relative weighting $10^{-3}$). 
Figure~\ref{fig:gz_results2}b shows that this weighting scheme favors GZ events from lighter element primaries with energies less than $10^{18}$ eV, and yields very slightly toward heavier primaries at the highest energies---a trend consistent with the observed cosmic ray spectrum~\cite{PDG}.
The observational consequence of this weighting scheme turning out to be opposite of reality would ultimately manifest itself in figure~\ref{fig:sigma3} as a left-translation (in minimum observation time) by an order of magnitude at most.

Additionally, the distribution of dual showers over the surface of the Earth is not isotropic, as it depends on the relative location of the Sun. A globally distributed sensor array which relies on consumer smartphones is likely to have a significantly higher effective area during the night, when phones are in darkness and not otherwise in use.   Figure~\ref{fig:gz_results3a} shows the geographically binned expected flux per year for examples of higher and lower energy cosmic rays.

It can be seen that there is generally a day/night asymmetry---the highest-energy GZ parent UHECRs shown ($10^{20}$ eV) are nearly 4-times more likely to dual-EAS on the sunny-side of Earth than the dark-side.
This asymmetry reverses, however, around $10^{17}$ eV where it becomes nearly 60-times more likely to dual-EAS on the dark-side of the Earth with $10^{15}$ eV UHECRs.
%

Unlike the flux distributions where the geographic bias changes with energy, the dark-side of the Earth consistently receives more closely-separated EASs than the sunny-side over all UHECR parent energies.  The time-of-day asymmetries in flux and EAS separation do not come so much as a surprise considering the probability to photodisintegrate is greatest for ``solar-passing'' trajectories (Fig.~\ref{fig:disint}) that experience highly Doppler-shifted (head-on) photons and high photon field density.
Solar-passing trajectories also experience the greatest HMF strengths, which tend to cause greater separations.
As the gyroradius (Eq.~\eqref{eq:Gyroradius}) for individual products scales in proportion to energy (and inversely with atomic number), one or both low-energy products increasingly misses the Earth (especially on the sunny-side of the Earth) resulting in the expected flux inversion favoring the dark-side of the Earth at low energies.

\begin{figure}[h!]
    \centering
    \includegraphics[width=\linewidth]{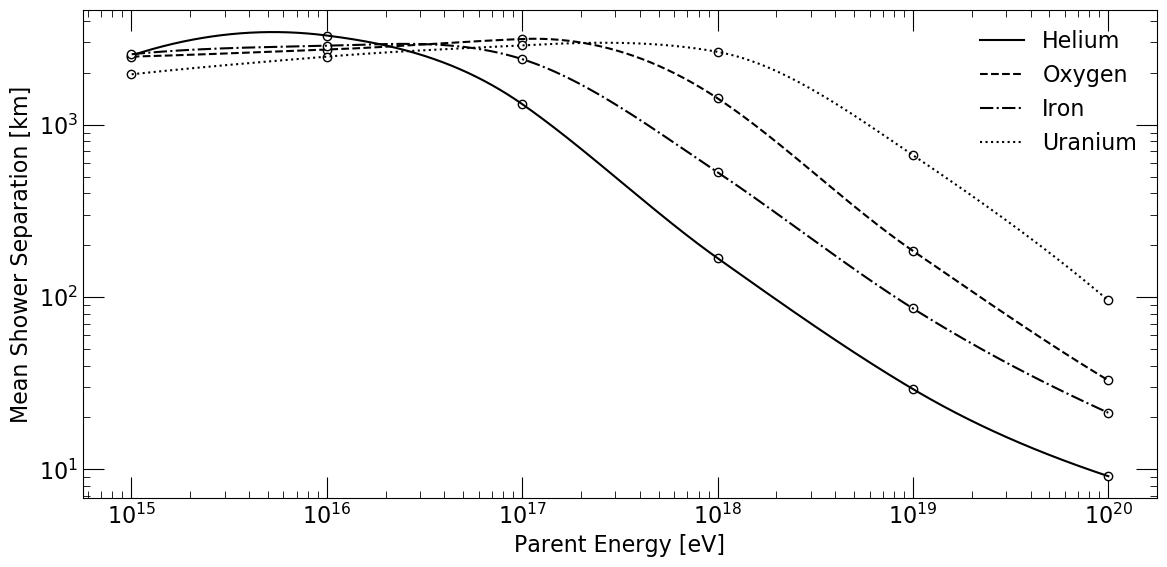}
    \caption{
      Mean separation between dual extended air showers caused by pairs of UHECRs generated by the GZ effect, a function of energy and primary atomic number.
    }
    \label{fig:gz_results4}
\end{figure}

On the other hand, although ``Earth-direct'' trajectories benefit from head-on (highly Doppler-shifted) photons, the photon field density is dramatically lower from this direction, resulting in photodisintegrations happening closer to the Earth on the average where the comparatively weak HMF (and shorter product propagation distance) results in more closely-separated EASs. 
The overall separation is greatest for the lowest energies, and smallest for the highest energy.
An average over all geographic bins is provided in Fig.~\ref{fig:gz_results4}.

\subsection{Expected rates of dual-EAS observations}

The expected rate of observed dual-EAS events in the CRAYFIS array under the two scenarios is the product of several factors: the rate of cosmic rays incident on the Earth, the probability for those particles to photodisintegrate and both decay products to strike the Earth, and the fraction of the resulting dual showers that are incident on CRAYFIS clusters of sufficient density.  Given an UHECR particle which disintegrates, the average energy of the lower-energy decay product is $\approx 0.23 E_{\textrm{UHECR}}$, given the energy splitting described in Eq.~\ref{eq:E_Split} under the imposed relative nuclei abundance in \S\ref{subsec:GZSimulations}.

\begin{figure}[h!]
    \centering
    \includegraphics[width=\linewidth]{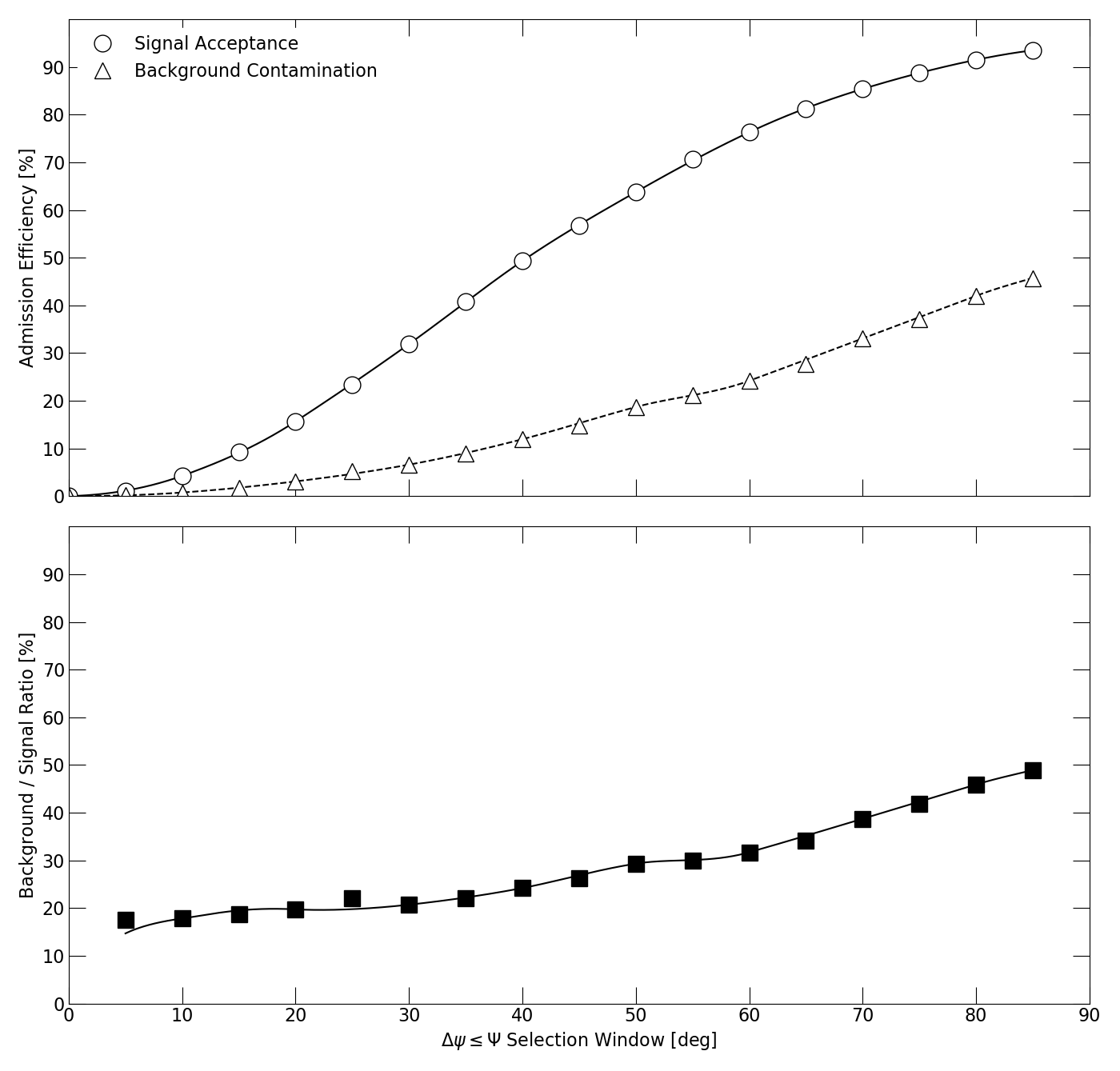}
    \caption{
        Top, expected signal acceptance (true $\Delta\psi = 0^{\circ}$) and background (true $\Delta\psi > 0^{\circ}$) contamination as a function of opening angle selection window limit, $\Delta\psi \leq \Psi$.
        The figure corresponds to a CRAYFIS angular resolution of ($\Delta\theta$, $\Delta\phi$) $=$ ($30^{\circ}$, $60^{\circ}$), where $\Psi=60^{\circ}$ corresponds to the efficiency of 80\%.
        The background follows a sinusoidal PDF in true $\Delta\psi$ (see Fig.~\ref{fig:bg_model}).
        Bottom, the ratio of background contamination to signal acceptance.
    }
    \label{fig:psi_window}
\end{figure}

In addition, we consider selection requirements to reduce the rate of the combinatoric background. Correlated dual EASs from the disintegration of a single nucleus will arrive along parallel trajectories, while randomly coincident showers will be more broadly distributed in angle.  Given an estimate of angular resolution of $\Delta\theta \approx 30^\circ$ and $\Delta\phi \approx 30^\circ$ from Ref.~\cite{crayfis}, we estimate the efficiency of a ceiling on $\Delta\Psi$, see Fig.~\ref{fig:psi_window}.

\begin{figure}[h!]
    \centering
    \includegraphics[width=\linewidth]{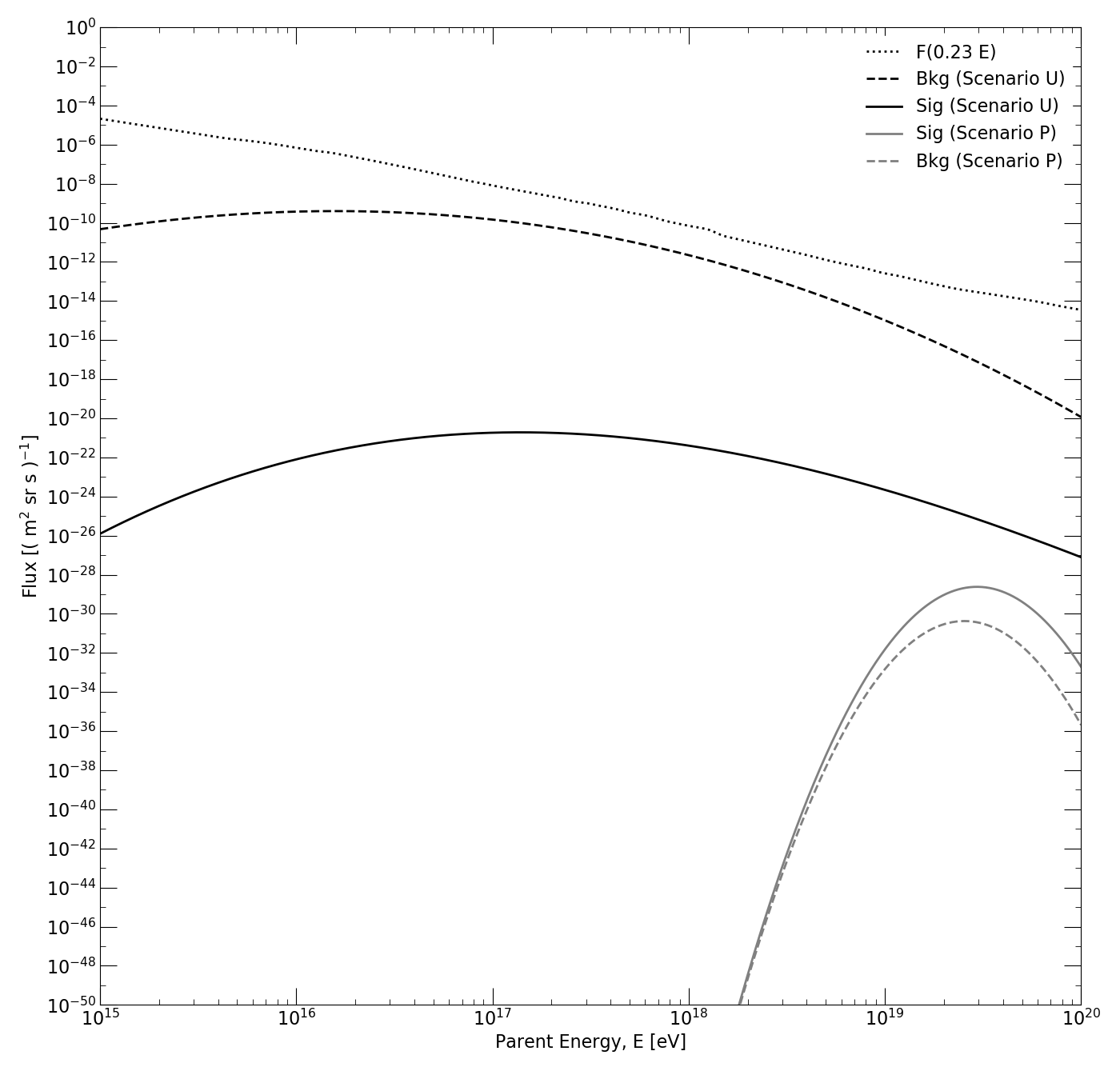}
    \caption{
        Expected GZ effect (and combinatorial background) dual-EAS flux within 100~ms time, and $60^{\circ}$ opening angle windows for ``Scenario U'' and ``Scenario P.''
        The established UHECR flux curve evaluated at the average nucleon EAS energy, $F(0.23\;E)$, is also shown as a comparison.
        See text for discussion.
    }
    \label{fig:money2}
\end{figure}

The background to signal rate drops, as expected, with increasingly tight ceilings on the opening angle. However, an overly restrictive selection will reduce sensitivity even as it enhances signal purity.  To balance these competing effects, we require $\Delta\Psi<60^\circ$.  The $\Delta\Psi$ selection efficiency is folded together with the cosmic ray flux so that the GZ probability and the efficiency to observe dual showers to estimate the dual-EAS flux in both scenarios, as well as the background rate, are shown in Fig.~\ref{fig:money2}.

\label{subsec:StatisticalAnalysis}
\subsection{Statistical Analysis}

The expected rate of dual EASs from the GZ effect is quite low, due to the small probability of the GZ effect itself and the difficulty of observing dual showers.  In scenario ``P,'' the rate of signal is many orders of magnitude below the observed flux. In scenario ``U'', the large number of observing sites leads to an amplification of random coincidences, swamping the signal enhancement; however, this is the result of considering the entire world-wide network at once.
Partitioning the globe into signal-optimized geographical regions of interest may show great improvement on background rejection.
In fact, scenario ``P'' partially demonstrates this by restricting events to only those that approach from the dark side of the Earth, and the limited number of sensitive land patches implicitly apply a cut on possible shower separations.
For scenario ``U,'' searches for events within fiducial bands of shower separation distances, primary energy, and geographic positioning relative the Sun are expected to provide powerful background suppressors; however this is left for future work.

We have used the GZ effect as a model for the expected signature of globally-coincident showers, but the larger question we address is whether a global array would be sensitive to any new, potentially unanticipated, source of simultaneous EAS.  The power of a new observatory is to discover the unexpected; new physics scenarios could boost the rate of GZ disintegration, or alternative models could provide showers with similar global distributions.  To address this broader question, we estimate the ability of the CRAYFIS array to observe coincident showers from some unknown process which creates GZ-like signatures at some higher rate.  

\begin{figure}[h!]
    \centering        
    \includegraphics[width=\linewidth]{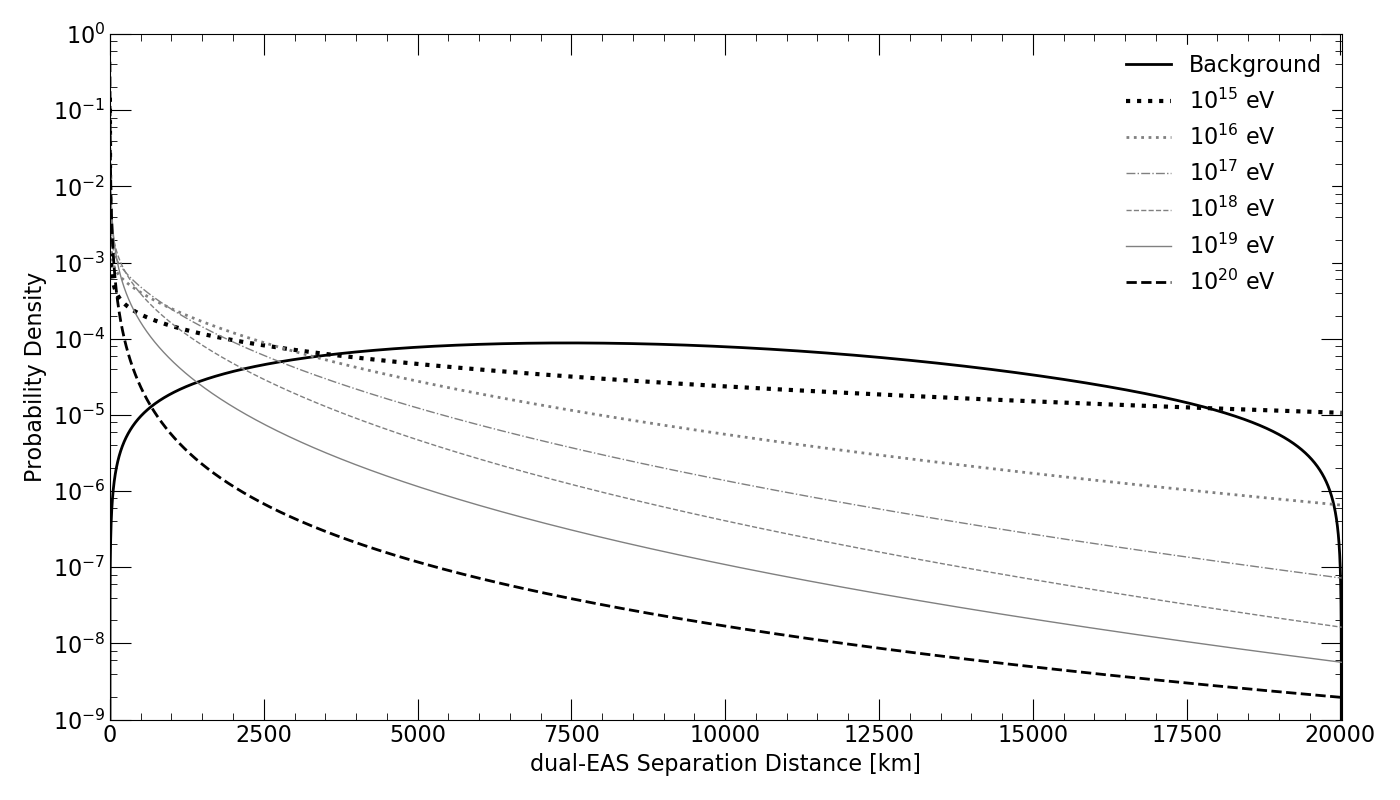}
    \caption{
        Top, the dual-EAS separation PDF for GZ effect-like signals at various characteristic energies (see legend), and the combinatorial background (thick, solid) (Eq.~\eqref{eq:background}) where $\Psi=60^{\circ}$.
    }
    \label{fig:sep_pdf}
\end{figure}

The appropriate statistical task is then to estimate the fraction of observed coincident showers which come from a generalized rate-boosted GZ-like process, rather than random background.  We have partially reduced the background rate by cutting on opening angle $\Delta\Psi$, but we have not yet taken advantage of the difference in expected distances $\Delta s$ between real and random coincidence. Figure ~\ref{fig:sep_pdf} shows this is a powerful differentiator, favoring true coincidences at smaller separations due to the reduced combinations for random coincidences.  The CRAYFIS sensitivity to GZ-like-effects that result in dual EASs is estimated using Monte Carlo pseudo-experiments with an unbinned likelihood in $\Delta s$.

\begin{figure}[h!]
    \centering
    \includegraphics[width=\linewidth]{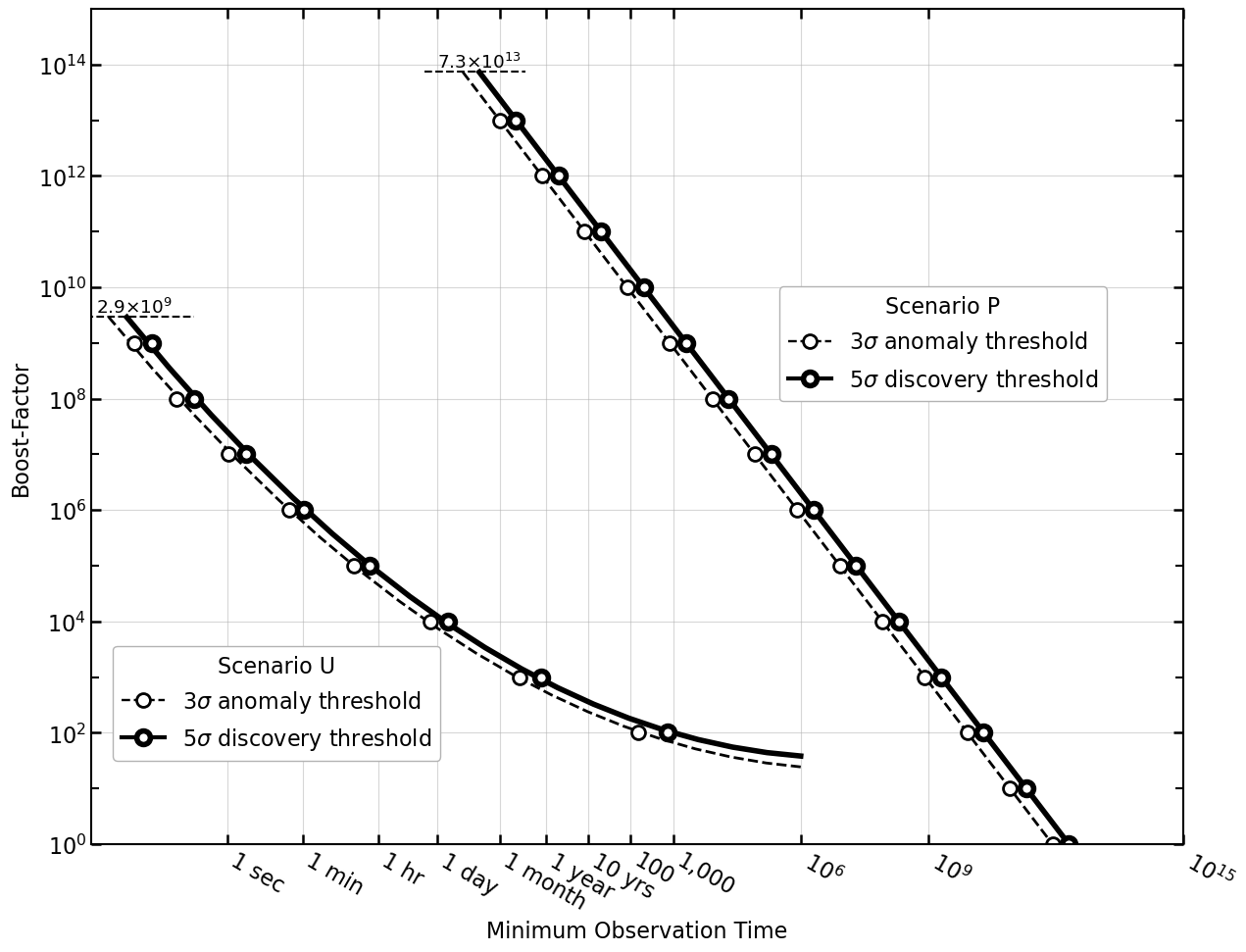}
    \caption{
        The minimum observation time needed to observe a GZ-like effect-like dual-EAS signal at  $3\sigma$ or $5\sigma$ significance as a function of the boost-factor, where boost factor of unity is the expected rate from GZ showers.    }
    \label{fig:sigma3}
\end{figure}

The separation distances for each of $N_\mathrm{tot}$ simulated dual-EAS events are drawn from the combined signal and background models (Fig.~\ref{fig:sep_pdf}), where $N_\mathrm{tot}$ is stepped by powers of 10 from $N_\mathrm{tot} = 10^{1}$~to~$10^{4}$, and for varying values of a rate-boosting factor, $B$, which simply increases the expected rate of showers relative to the GZ effect.
The boost-factors for ``Scenario U'' range in powers of 10 from $B_\mathrm{U} = 10^{0}$~to~$10^{9}$, and for ``Scenario P'' from $B_\mathrm{P} = 10^{0}$~to~$10^{13}$---where the greater boost is capped to be consistent with the established UHECR flux.
The minimal observation time to reject the background-only hypothesis at $3\sigma$ significance is shown in Fig.~\ref{fig:sigma3}.

\section{Discussion}

In an ideal upper-limit scenario where a CRAYFIS array covers the Earth (``Scenario U,''), such an array is capable of statistically significant discovery of spatially-separated but coincidence showers in reasonable short operation periods ($<1$ year) if such showers are produced at rates at least $10^3$ times the expected rate from the GZ effect.

In the more realistic scenario (``Scenario P,'') in which a CRAYFIS array operates part-time on approximately $10^6$ devices, much higher rates of dual-shower production ($>10^{12}$) are required for statistically significant observations. For this reason, smaller CRAYFIS array scenarios are probably only likely (under reasonable time frames) to detect ``burst'' phenomena where many simultaneous EASs occur at once, versus ``continuous'' phenomena like the GZ effect.

It should be remembered, however, that the GZ effect is just one such phenomenon described here as a concrete benchmark.  Given our near-total lack of observational power to detect simultaneous showers, the true rate of such events is essentially unknown. A network of smartphones operating together could provide us with our first glimpse into phenomena with Earth-scale correlations.

\label{sec:Discussion}

\section{Acknowledgments}
\label{sec:Acknowledgements}

DW is supported by the Department of Energy Office of Science.
EA is supported by NSF fellowship (NRT-DESE) award number 1633631.
This work was performed under the auspices of the U.S. Department of Energy by Lawrence Livermore National Laboratory under Contract DE-AC52-07NA27344.


\appendix


\begin{table}
\caption{Important parameter values for the CORSIKA simulation}
\label{tab:corsika}
\begin{tabular}{lr}
\hline\hline
Parameter & Value\\
\hline
ECUTS & 0.05 GeV (hadrons)\\
& 0.01 (muons)\\
& 0.00005 (electrons)\\
& 0.00005 (photons)\\
ECTMAP & 0.001\\
THETAP & 0 \\
PHIP & 0 \\
OBSLEV & 0, 500E2, 1000E2, 1400E2 \\
& 2000E2, 5000E2, 10000E2 \\
& 20000E2, 50000E2, 100000E2\\
FIXCHI & 0\\
ATMOD & 1 \\
MAGNET & 21.82 45.51\\
THIN & 1.E-6 1.E30 0.E0\\
ELMFLG & T\ T \\
RADNKG & 20000E2 \\
STEPFC & 1.\\
MUMULT & T \\
HADFLG & 0\ 1\ 0\ 1\ 0\ 2 \\
\hline\hline
\end{tabular}
\end{table}


\raggedright
\Urlmuskip=0mu plus 1mu\relax
\bibliographystyle{aasjournal}
\bibliography{paper}

\end{document}